\documentclass{emulateapj-rtx4}
\usepackage{amssymb, amsmath, epsfig} 

\newcommand{\grad}{{\mathbf{\nabla}}}

\newcommand{\calM}{{\cal M}}
\newcommand{\calMbc}{{{\cal M}_{\rm bc}}}

\newcommand{\bfk}{{\boldsymbol{k}}}

\newcommand{\bfv}{{\boldsymbol{v}}}
\newcommand{\bfq}{{\boldsymbol{q}}}
\newcommand{\bc}{{\rm bc}}

\newcommand{\bfx}{{\boldsymbol x}}

\newcommand{\cMpc}{{\rm Mpc}}
\newcommand{\ckpc}{{\rm kpc}}

\newcommand{\Msun}{{M_{\odot}}}

\newcommand{\vbc}{{v_{\rm bc}}}

\newcommand{\change}[1]{{#1}}

\def\apjl{ApJL}
\def\apjs{ApJS}

\def\aap{{A\&A}}

\begin{document}

\title{The formation of the first cosmic structures and the physics of the $z\sim 20$ Universe.}
\author{ Ryan M.\ O'Leary\altaffilmark{1}$^,$\altaffilmark{2} \& Matthew McQuinn\altaffilmark{1}$^,$\altaffilmark{2}}

\altaffiltext{1} {Einstein Fellows\\}
\altaffiltext{2} {Department of Astronomy, University of California, Berkeley, CA 94720, USA\\}

\begin{abstract}

We perform a suite of cosmological simulations in the $\Lambda$CDM paradigm of
the formation of the first structures in the Universe prior to astrophysical reheating and reionization ($15 \lesssim z < 200$).  These are the first simulations initialized in a manner that self consistently accounts for 
the impact of pressure on the rate of growth of modes, temperature
fluctuations in the gas, and the dark matter--baryon supersonic velocity
difference.  
Even with these improvements, these are still difficult times to simulate accurately as the Jeans length of the cold intergalactic gas must be resolved while also capturing a representative sample of the Universe.  We explore the box size and resolution requirements to meet these competing objectives.

Our simulations support the finding of recent studies that the dark matter--baryon velocity difference has a surprisingly
large impact on the accretion of gas onto the first star-forming
minihalos (which have masses of $\sim 10^6~\Msun$).  In fact, the halo gas is often significantly downwind of such halos and with lower densities in the simulations in which the baryons have a bulk flow with respect to the dark matter, modulating the formation of the first stars by the local value of this velocity difference.  We also show that dynamical friction plays an important role in the nonlinear evolution of the dark matter--baryon differential velocity, acting to erase this velocity difference quickly in overdense gas as well as sourcing visually-apparent bow shocks and Mach cones throughout the Universe.  

 We use simulations with both the GADGET and
Enzo cosmological codes to test the robustness of these conclusions.  The comparison of these codes' simulations also provides a relatively controlled test of these codes themselves, allowing us to quantify some of the tradeoffs between the algorithms.  For example, we find that particle coupling in GADGET between the gas and dark matter particles results in spurious growth that mimics nonlinear growth in the matter power spectrum for standard initial setups. \change{This coupling is alleviated by using adaptive gravitational softening for the gas.}
 In a companion paper, we use the simulations presented here to make detailed estimates for the impact of the dark matter--baryon velocity differential on redshifted 21cm radiation.  The initial conditions generator used in this study {\sc CICsASS} can be publicly downloaded. 
\end{abstract}


\section{introduction}
For the first hundred million years after the Big Bang, the gas distribution in the Universe as well as its thermal state can be accurately calculated by solving a set of linear differential equations.
However, between the redshifts of $100$ and $10$, the inhomogeneties in much of the cosmic gas went nonlinear.  Eventually, 
deep enough potential wells for the primordial gas to cool and form stars developed,
and the Universe transitioned to a vastly
more complex system in which stars abound and their radiative backgrounds impacted all of the baryonic matter.  In principle, it is
possible to understand perfectly the evolution of gas and dark matter before stellar radiation backgrounds impacted all matter, at $z \gtrsim 20 $, using a combination of
linear theory and (once nonlinear structure begins to form) numerical simulations. Nevertheless, important
questions about the evolution of the Universe during this virgin epoch
remain unanswered.

For example, it is unclear whether weak structure formation shocks
would have significantly heated the cosmic gas \citep{gnedin04,furlanettoshocks}.  In the absence of shocks, the intergalactic medium (IGM) is anticipated
to have been kinetically cold prior to reheating by astrophysical
sources, reaching $10~$K at $z=20$, and with the temperature cooling adiabatically as
$(1+z)^2$.  However, even $0.3~$km~s$^{-1}$ flows would have been
supersonic for a gas temperature of $10~$K, and supersonic motion is
likely to source shocks and entropy generation.

In addition, recently \citet{tseliakhovich10} demonstrated that in
most places in the early Universe the baryons and dark matter were
moving supersonically with respect to one another.  At the time of
recombination, the cosmic gas was moving with respect to the dark
matter at an RMS velocity of $30~$km~s$^{-1}$ and in a coherent manner
on $\lesssim10$~comoving~Mpc separations.  These initial velocity
differences translate into the dark matter moving through the gas
with an average Mach number of $\calMbc \approx 1.7$ over $15 \lesssim z <
150$, but with a standard deviation in Mach number between different regions in the Universe of $0.7$.  As with structure formation, such supersonic motion may source shocks, generating entropy and reheating the universe.

In this paper, we simulate the evolution of matter in the
Universe prior to when radiation backgrounds generated by stars became important sources of heating.  This time has been the focus of many prior studies of the first stars (e.g., \citealt{abel02, bromm02}).  We discuss the simulation box size and resolution requirements to simulate these times accurately, and we add several improvements to standard methods for initializing cosmological simulations so that our simulations are initialized with full linear solutions for the growth of structure.  For example, in contrast to prior studies, our initial conditions self-consistently account for the impact of gas pressure on the growth of modes as well as include fluctuations in the gas temperature (an improvement emphasized as important in \citealt{naoz05} and \citet{naoz07}).  

A few prior studies have investigated the impact of the supersonic motion of the gas relative to the dark matter on the formation of the first gas-rich halos and on the first stars \citep{maio11, stacy11, greif11, naoz11b}.    Interestingly, some of these studies find this motion has a dramatic impact on the formation of the first stars.  However, the relative velocity in all of these simulations was incorporated by boosting the velocity of the gas at the onset of the simulation, which we show misses much of the impact of this supersonic motion on the linear growth of structure.   The simulations in this study are the first to use a consistent linear theory to initialize these differential flows.  Our simulations enable us to more rigorously test these claims as well as to investigate other manifestations of such cosmic flows.

The Universe during the `Dark Ages' -- times before
stars reionized and reheated the Universe -- is observable via the
redshifted 21cm line in absorption against the cosmic microwave
background.  Several collaborations are currently developing
instruments to detect this era (LEDA, DARE, and LOFAR\footnote{\url{www.cfa.harvard.edu/LEDA}, \citep{burns11}; \url{http://lunar.colorado.edu/dare/}, \citep{bernardi12}; \url{http://www.lofar.org/}, \citep{harker10}}).  The strength
of the signal, especially on large scales, is intimately related the
both thermal history of the gas as well as the star formation rate
\citep{madau97, furlanetto06, furlanettoohbriggs}.  In a companion paper (\citealt{paperII}; hereafter Paper II), we will address
the observational signatures of this era in the redshifted $21$cm
absorption signal, specifically focusing on the impact of the
relative velocity between the baryons and dark matter.

This paper is organized as follows:  Section \ref{sec:scales} elucidates the characteristic scales and physical processes that affect the evolution of intergalactic gas at $15 \lesssim z < 200$.  Section \ref{sec:sims} discusses considerations relevant to simulating these early cosmic times as well as the details of our initial conditions generator.  This section also compares  cosmological simulations of the early Universe run using both the GADGET \citep{springel01} and Enzo \citep{oshea04} codes.  Section \ref{ss:df} describes the important roll of dynamical friction in the non-linear evolution of structure formation with a baryonic streaming velocity, $\vbc$.  Section \ref{sec:results2} uses these simulations to characterize the properties of this cosmic epoch as well as the impact of $\vbc$ on structure formation.  This study assumes a flat $\Lambda$CDM cosmological model with $\Omega_m=0.27$, $\Omega_\Lambda=0.73$, $h=0.71$, $\sigma_8= 0.8$, $n_s=0.96$, $Y_{\rm He} = 0.24$, and $\Omega_b = 0.046$, consistent with recent measurements \citep{larson11}.  We define the $z=0$ dark matter fraction as $\Omega_c \equiv \Omega_m -\Omega_b$.  We will subsequently abbreviate proper Mpc as pMpc, and Mpc and kpc will be reserved for comoving lengths.  Some of our calculations use the Sheth-Tormen mass function, for which we adopt the parameters $p = 0.3$, $a=0.75$, $A=0.322$  \citep{sheth02}.\\

\section{Characteristic Scales in the Post-Recombination and Pre-Reionization Universe}
\label{sec:scales}

The focus of this paper and Paper II is on the era after the gas
thermally
decoupled from the CMB and before it was reheated by astrophysical
sources.  This period is anticipated to have occurred over $15
\lesssim z \lesssim 200$ \citep{peeblesbook, gnedin04, furlanetto06}.
It should be possible to model most aspects of these pristine times
from the cosmological initial conditions alone.  The temperature of
the gas at the cosmic mean density cooled adiabatically with the
expansion of the Universe during this period, decreasing with redshift
as $(1+z)^2$ and reaching a temperature of $10~$K at $z=20$.  In fact,
the vast majority of the gas likely existed near a single adiabat:  It is unclear when or if
structure formation shocks would have contributed significantly to the
entropy of the IGM (a question investigated here), and only
in the most massive, rarest halos was the gas able to cool by
molecular or atomic transitions.  In addition, only at the end of
this adiabatic period was more than a percent of the gas bound to dark
matter halos.  In particular, the fraction of matter that had
collapsed into dark matter halos with masses $ > 3\times 10^4~\Msun$
-- those massive enough to overcome pressure and retain gas
\citep{naoz09} -- at $z= 15, 20,$ and $30$ was $0.03,~ 0.005$, and
$6\times10^{-5}$, according to the Sheth-Tormen mass function.  These
numbers become $0.008,~ 7\times10^{-4}$, and $1\times10^{-6}$ for $>
10^6\Msun$ halos -- halos massive enough to cool via molecular
hydrogen transitions and host stars \citep{tegmark97}.

\begin{figure}
\includegraphics[width=\columnwidth]{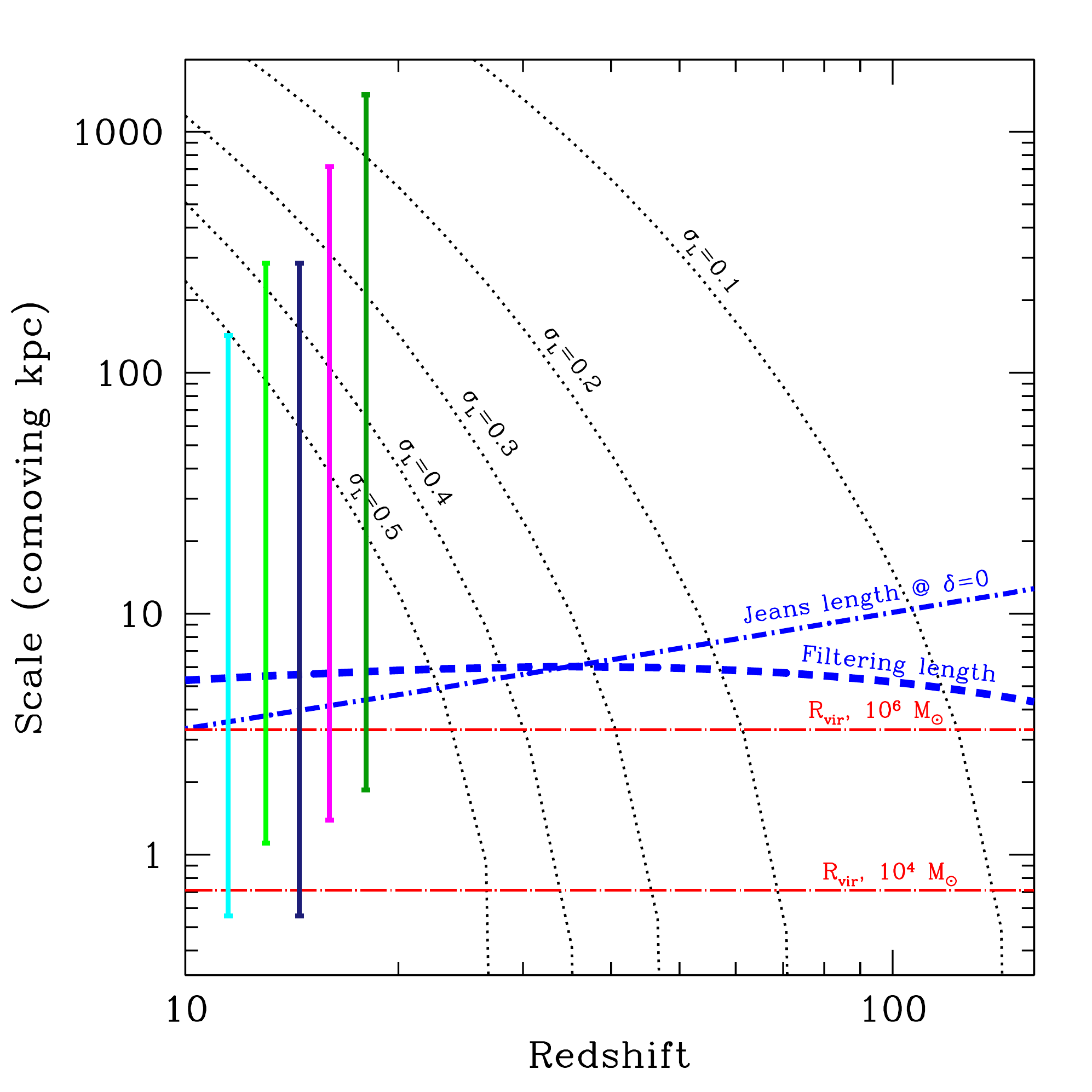}
\caption{Characteristic comoving scales in the Universe prior to astrophysical reheating: the virial radius of $10^{4}~\Msun$ and $10^{6}~\Msun$ halos, the Jeans' length for mean density gas, the Filtering length, and the radius of a sphere with RMS linear density fluctuation $\sigma_L$ of the specified value.  Each vertical bar represents one of the simulations employed in this study, with its height stretching from the mean interparticle distance (or base-grid mesh width) to the simulation's box size.  The leftmost bar represents a simulation with $\{0.1/h~\cMpc, ~256^3~{\rm gaseous ~resolution~ elements}\}$, and the rightmost one with $\{1/h~\cMpc,~768^3~{\rm gaseous ~resolution~ elements}\}$.   \label{fig:char_scales}}
\end{figure}

Figure \ref{fig:char_scales} shows the scales most relevant to the said epoch in comoving coordinates:  the virial radius of a $10^4~\Msun$ halo and of one with $10^6~\Msun$, the Jeans' and Filtering length as defined in \citet{naoz07}, as well as the radius of a sphere in which the RMS linear density contrast, $\sigma_L$, equals $0.1$, $0.2$, $0.3$, $0.4$, and $0.5$.  The virial radii of halos that can retain their gas and form stars, the Jeans' length, and the Filtering length are all $\sim 1-10~$\ckpc.   In fact, the comoving Jeans' length, $R_J$, is nearly constant over $15 \lesssim z < 200$, and equal to $5~$\ckpc~ with the scaling $R_J \propto (1+\delta)^{1/6} (1+z)^{1/2}$, where $\delta$ is the matter overdensity.  The comoving Filtering length, which is the analogue of the Jeans' length at the mean density of an expanding universe \citep{gnedin98, naoz07}, also has a weak dependence on redshift.\footnote{
Interestingly, \citet{naoz09} found that the linear-theory Filtering mass characterizes the minimum mass halo that can overcome pressure and retain its gas; smaller mass halos are largely devoid of gas.  We think there is a simple explanation for why the Filtering mass sets the minimum mass of a halo that can retain its gas.  For adiabatic collapse (appropriate for almost all gas at the specified epoch), the Jeans' mass \emph{increases} with density as $\rho^{1/2}$.  In addition, for gas in the Hubble flow, the characteristic scale above which gas fragments is the Filtering mass.  For realistic thermal histories, the Filtering mass is even smaller than the Jeans' mass, at least at the time when a region that virializes at $z\sim 20$ decoupled from the Hubble flow.  Therefore, collapsing gas formed its smallest unit when it was in the Hubble flow, and these units were unable to fragment further as they collapsed and instead were dragged into the dark matter potential wells as they formed.  Following the same logic, the filtering mass should be a poor approximation for the characteristic halo mass that is able to retain its gas after reionization.  In this case, the gas was photoheated to $~10^4~$K by reionization prior to collapse.   This initial temperature floor results in the collapse being non-adiabatic and instead being better approximated as isothermal once the gas had collapsed to moderate overdensities such that it has adiabatically heated to $\sim 2-3 \times 10^4~$K (temperatures where collisional cooling is extremely efficient).   For isothermal gas, the Jeans' mass scales as $\rho^{-1/2}$; collapsing regions fragment into smaller and smaller clumps as their density increases.  In support of the hypothesis, \citet{hoeft06} and \citet{okamoto08} find using numerical simulations that this characteristic halo mass that retains gas after reionization is generally substantially smaller than the Filtering mass.  In addition, this also provides an explanation for why studies using cosmological simulations find that the sizes of overdense gaseous clumps in the photoionized IGM are approximately set by the Jeans' scale for the clump's current density rather than the Jeans' scale at the cosmic mean density \citep{schaye01, mcquinn11, altay11}.}   The vertical line segments in Figure \ref{fig:char_scales} show the range of scales that are captured by the simulations we run for this study and discussed in \S \ref{sec:sims}.

\section{Simulating the Dark Ages}
\label{sec:sims}

\subsection{Simulation Parameters and Initial Conditions}
\label{ss:ics}

\begin{figure}
\rotatebox{-90}{
\includegraphics[height=11cm]{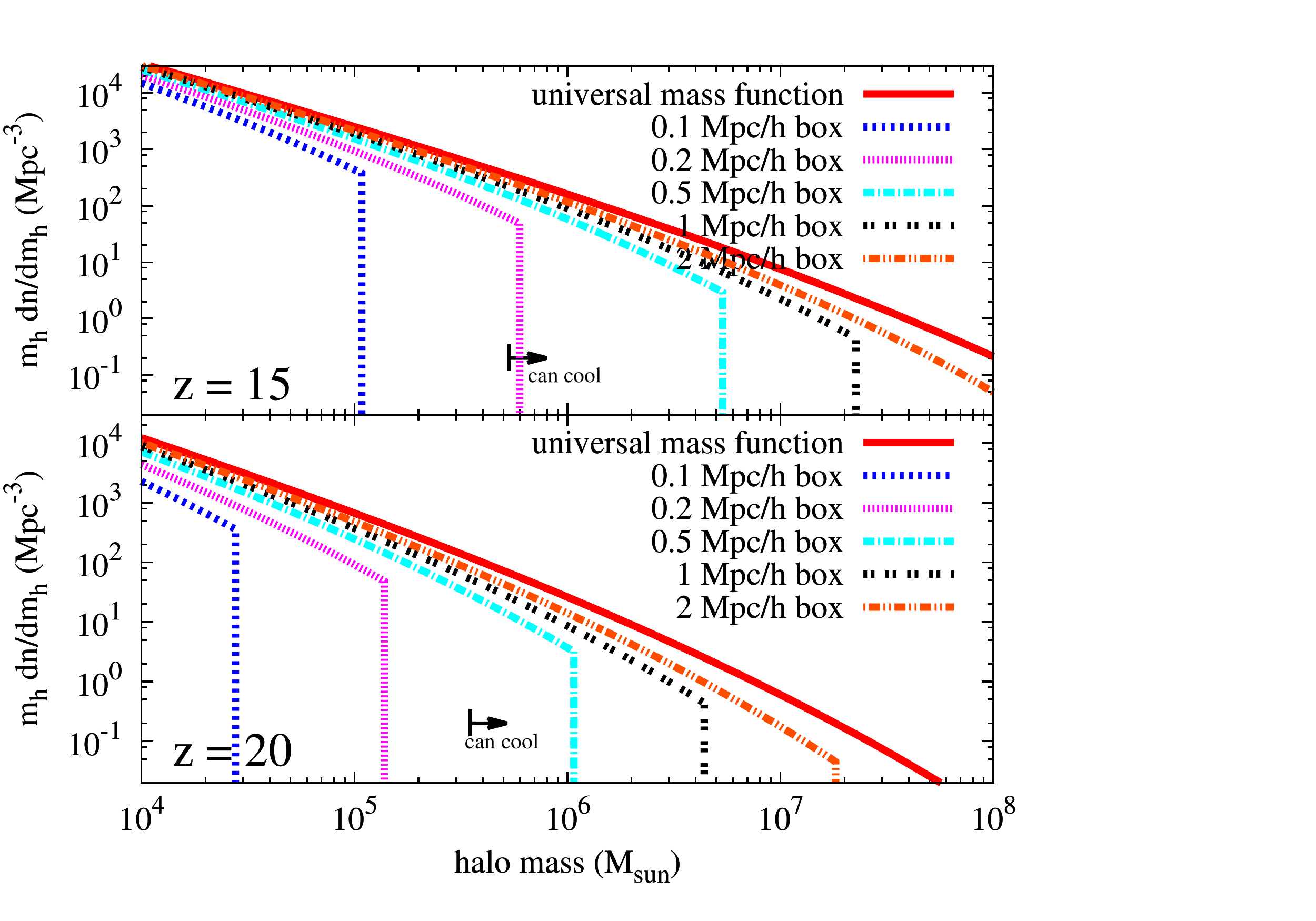}}
\caption{Predicted halo mass function, $dn/dm_h$, for simulations with box sizes that span the range studied here (as well as a hypothetical $2~\cMpc/h$ box).   Shown is the expected mass function, which we set to zero when $m_h dn/dm_h \, L_{\rm box}^3 < 1$.  The rightward pointing arrow indicates the halo masses at which inter-halo gas has the potential to cool via molecular emissions and form stars (using the criterion $v_{\rm cir} > 3.7~$km~s$^{-1}$; e.g., \citealt{fialkov11}).  \label{fig:mf}}
\end{figure}

There are a few hurdles that must be overcome in order to accurately simulate the formation of the first structures in the Universe and the impact of the dark matter--baryon velocity differential:\\ 

First, a simulation of these times needs to capture the pressure smoothing scale of the gas (or $\sim 5$ comoving kpc between $20 \lesssim z \lesssim 200$; Fig. \ref{fig:char_scales}) while also being large enough so that the box--scale modes are still linear.  Otherwise, it does not properly capture the gas physics and/or is not representative of the Universe.   As illustrated in Fig.~\ref{fig:char_scales}, capturing a representative volume becomes increasingly difficult with decreasing redshift around $z\sim 20$ because the nonlinear scale rapidly moves to larger scales with decreasing redshift owing to the near scale-invariance of Mpc-scale density fluctuations.  A non-representative box size has a dramatic effect on the halo mass function \citep{barkana04}.  Figure \ref{fig:mf} shows the range of halo masses that different simulation box sizes capture.\footnote{The halo mass function as a function of box size in Figure~\ref{fig:mf} is calculated by multiplying the Sheth-Tormen mass function by $n_{\rm PS}(m_h | \sqrt{\sigma_{m_h}^2 - \sigma_{l_{\rm box}/2}^2})/ n_{\rm PS}(m_h | \sigma_m)$, where $n_{\rm PS} (m_h, \sigma_X)$ is the Press-Schechter mass function at mass $m_h$ given $\sigma_X$, the RMS density contrast in a sphere of radius $X$.  This prescription was motivated in \citet{barkana04}.}  A $1~$Mpc$/h$ box is needed to statistically capture the $z=20$ halo mass function at the factor of $\lesssim 2$--level at the mass threshold that can cool by molecular hydrogen transitions, and an even larger box size is required to meet this requirement to higher redshifts.

\begin{figure}
\begin{center}
\rotatebox{0}{
\includegraphics[height=\columnwidth]{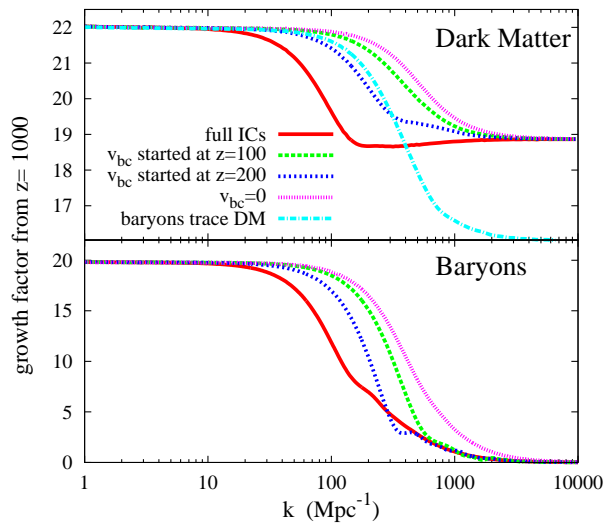}}
\end{center}
\caption{Investigation of different approximations for initializing numerical simulations with $v_{\rm bc}$.  Each curve is the linear growth factor at $z=30$ of the dark matter (top panel) and the gas (bottom panel), initialized so that it equals unity at $z=1000$, and for modes with $v_{\rm bc} \cos \phi_k = 3$ km~s$^{-1}$ at $z=100$ ($\calM_{\rm bc} \cos \phi_k = 1.8$) unless labelled otherwise.  All previous simulations added a relative dark matter--baryon velocity on top of the standard cosmological initial conditions at the initialization redshift of the simulation, either $z=100$ or $z=200$.  As can be gauged from comparing to the sold curve, which represent the full solution as discussed in \citet{tseliakhovich10}, to the long and short dashed curves, such an initialization misses much of the effect of $v_{\rm bc}$ in linear theory.  In addition, previous simulations had not included pressure self-consistently in the initial conditions.  If the baryons are initialized at $z=100$ with $\calM_{\rm bc} = 1.8$ and such that they trace the dark matter distribution as in most previous studies, for $\cos \phi_k = 1$ this leads to the cyan dot-dashed curve in the top panel. \label{fig:GF}}
\end{figure}

Second, the effects of the baryonic streaming velocity on the growth of Jeans'-scale perturbations is important over a broad range of redshifts after the baryons thermally decoupled from the CMB \citep{tseliakhovich10}.
\change{Figure \ref{fig:GF} shows solutions to the linearized equations (\ref{eqn:lintheory1} - \ref{eqn:lintheory4} in Appendix \ref{ap:linth}) using initializations of $v_{\rm bc}$ common in the numerical literature as well as the full linear theory initialization used in this paper.  Each curve is the linear growth factor at $z=30$ of the dark matter (top panel) and the gas (bottom panel), initialized so that it equals unity at $z=1000$.  As can be gauged from comparing to the sold curve (which represent the initial conditions used in this study where $v_\bc$ is consistently followed) to the long and short dashed curves (which approximate the initial conditions used in \citealt{maio11}, \citealt{stacy11}, \citealt{greif11}, and \citealt{naoz11b} where $v_\bc$ is incorporated only at the onset of the simulations), simply boosting the velocity of the baryons at $z=100$ or $200$ misses much of the linear effect on $\delta_b$ and $\delta_c$.  The dotted curve is the case with $v_{\rm bc} = 0$.  In addition, the vast majority of studies have assumed that the baryons also trace the dark matter, which results in the baryons streaming out of the potential wells of the dark matter early on in the simulations and leads to the dot-dashed curve in the top panel.}

Lastly, simulations exploring the high-redshift Universe cannot be
initialized at similar redshifts to those that are used to understand
lower redshifts.  A halo that collapses at $z=25$ has a linear
overdensity of $0.44$ at $z=100$ in the spherical collapse model
\citep{gunn72} and so the error in the overdensity from using only Eularian (Lagrangian) linear theory
is uncomfortably large, $52\%$ ($11\%$), with this
error resulting in the structures being less bound and collapsing at later times.  See
\citet{crocce06} for more quantitative determinations of this error.
Initial conditions that use $2^{\rm nd}$~order perturbation theory would reduce this error, but higher order
solutions that account for gas pressure have only been developed for
toy cases \citep{shoji09}.  At the same time, the particle noise in
simulations initialized at too high of a redshift dominates over the
cosmological clustering.  Thus, a balance must be reached between
initializing at a redshift where $1^{\rm st}$~order Lagrangian
perturbation theory is accurate and avoiding particle noise.  Prior
studies of $v_{\rm bc}$ (as well as most studies of the first stars;
\citealt{abel02, greif11}) typically used $z_i\approx100$ and $1^{\rm
  st}$ order Lagrangian perturbation theory.  We favor higher
redshifts in this study, with $z_i\approx200$ or $400$, since this choice does not significantly increase the computational
requirements of the simulations nor add much extra noise to the power spectrum of the density field (\S \ref{ss:nonlin}).

To initialize our simulations, we adopt the approximation in \citet{tseliakhovich10} (and further motivated in \S~5.3.1 in \citealt{hu95} and \citealt{eisenstein98}) that the baryons decoupled instantaneously at $z=1000$ and solve equations (\ref{eqn:lintheory1}-\ref{eqn:lintheory4}) for the growth of matter fluctuations in linear theory to redshift $z_i$.  These $1^{\rm st}$ order differential equations are initialized with the CAMB Boltzmann code transfer function\footnote{\url{http://camb.info/}}  and its time derivative. We find that this approximation reproduces the evolution of the matter power in CAMB excellently for $v_\bc = 0$.  To relate the linear theory solution to particle displacements, we use that the linearized displacement $\mathbf{\Psi}$ from Lagrangian position $q$ is related to the Eularian linear perturbation theory overdensity, $\delta_E$, via the relationship $\grad \cdot \mathbf{\Psi} = - \delta_E$ for an irrotational flow (the ``Zel'dovich approximation''; e.g., \citealt{padmanabhan}).  It follows that the linear displacement  and its velocity are given by
\begin{eqnarray}
\bfx &\equiv& \bfq + \mathbf{\Psi} =   \bfq - \grad \phi,\label{eqndisp}\\
\bfv &=&  - a\,  \grad \dot{\phi},
\label{eqnvel}
\end{eqnarray}
where $\grad^2 \phi = \delta_E$.  These equations set the displacements that are used from the initial positions set by a glass file (using cloud--in--cell interpolation from a grid with each dimension equal to the cube root of the particle number).\footnote{Linear order Lagrangian perturbation theory (the Zel'dovich approximation) \change{smooths out small scales, resulting} in the dark matter power spectrum at lowest order being given by 
\begin{equation}
P_Z(\bfk) = P_E (\bfk) \exp[-\sigma_R^2 \,k^2/2],
\end{equation}
where $P_E$ is the linear theory Eularian density power and $\sigma_R^2 = (3 \pi)^{-2}\int_{\rm k_{\rm min}}^{\rm k_{\rm max}} dk \,P_E(k)$.  Thus, this theory will be valid at $k \ll \sigma_R^{-1}$ such that $P_Z(\bfk) = P_E$.  In a $1~$Mpc box, $\sigma_R = 0.3 \, G(z)~$Mpc, where $G(z)$ is the growth factor.  To illustrate this difficulty, for a $0.1~$Mpc box, $k_* = \sigma_R^{-1} = 3400~$Mpc$^{-1}$ at $z=200$, which is comparable to the Nyquist frequency,  $k_N = 8000~ (N/256)~\cMpc^{-1}$.  Here, $N$ is the cube root of the particle number.  For a $0.5~$Mpc box, $k_*$ becomes $k_* = 900~$Mpc$^{-1}$ and for a $1~$Mpc box, $k_* = 500~$Mpc$^{-1}$.   
These numbers may be prohibitive if we were attempting to capture the smallest scales (and smallest dark matter halos) in our box.  However, $k_N$ is a scale that is generally buried in the particle noise.  The criterion that needs to be satisfied is $k_* \gg k_F$, where the Filtering wavevector $k_F$ is $\sim 500~\cMpc^{-1}$.  In all of the simulations employed here, this condition is at least weakly satisfied.  Zoom in simulations of the first stars, where $k_{\rm min}$ can be much smaller than in our simulations, should be wary of this deficiency of $1^{\rm st}$ order Lagrangian perturbation theory.}  For our SPH simulations, we typically also use a different glass file to initialize the baryons as recommended in \citet{yoshida03} to avoid spurious couplings between different particle types, \change{although we found this resulted in increased coupling in some simulations} (see Appendix B).

\begin{figure*}
\begin{center}
\includegraphics[width=5cm]{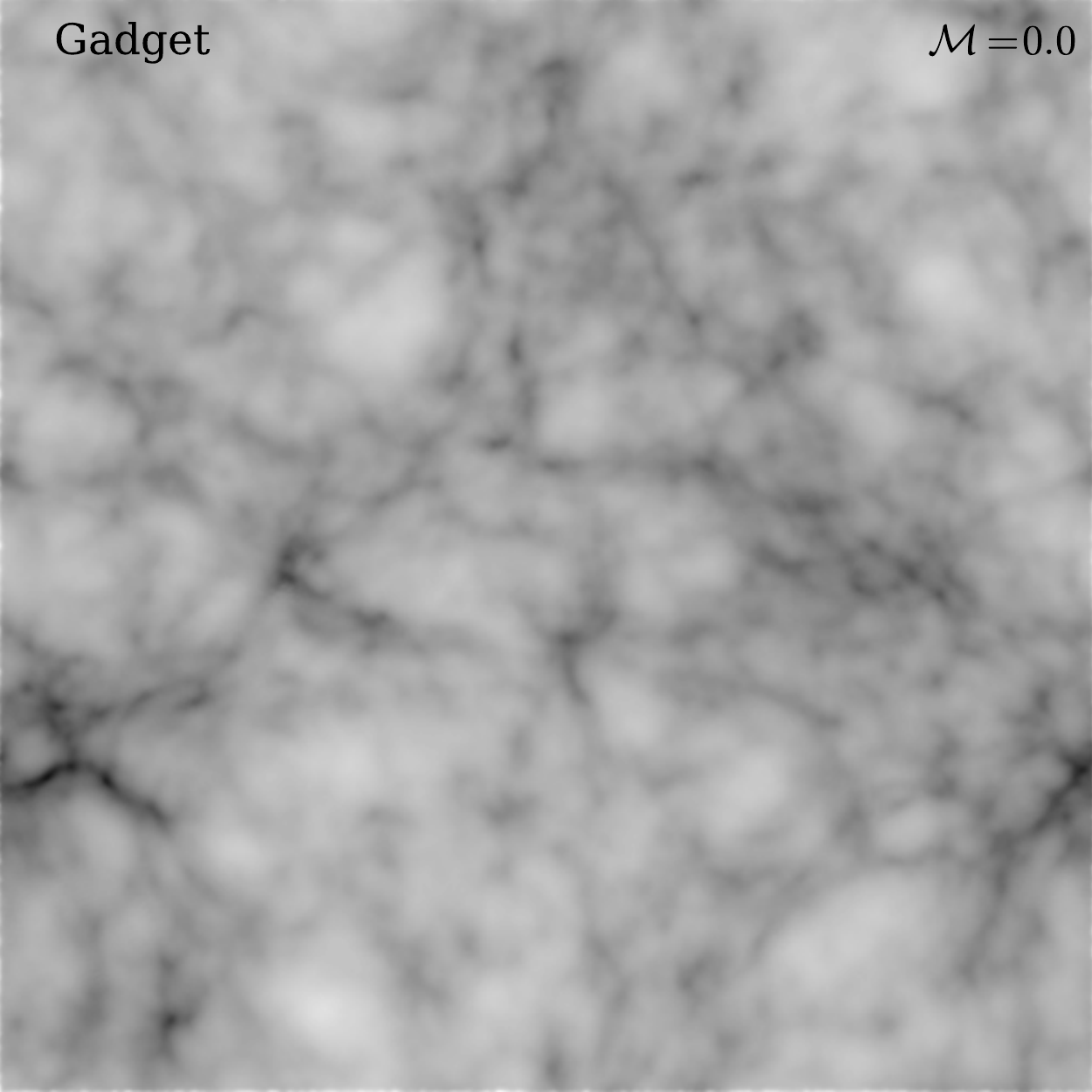}
\includegraphics[width=5cm]{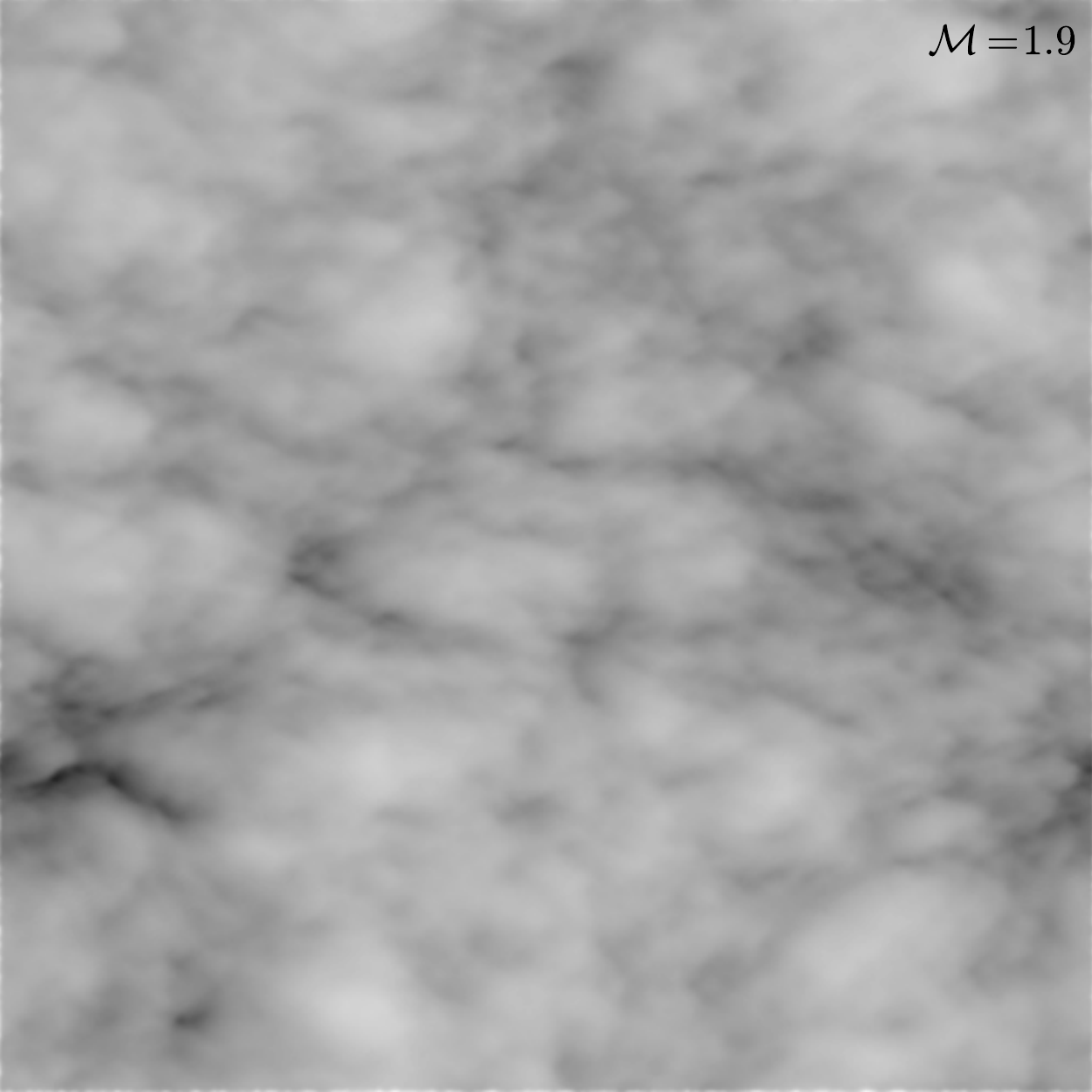}
\includegraphics[width=5cm]{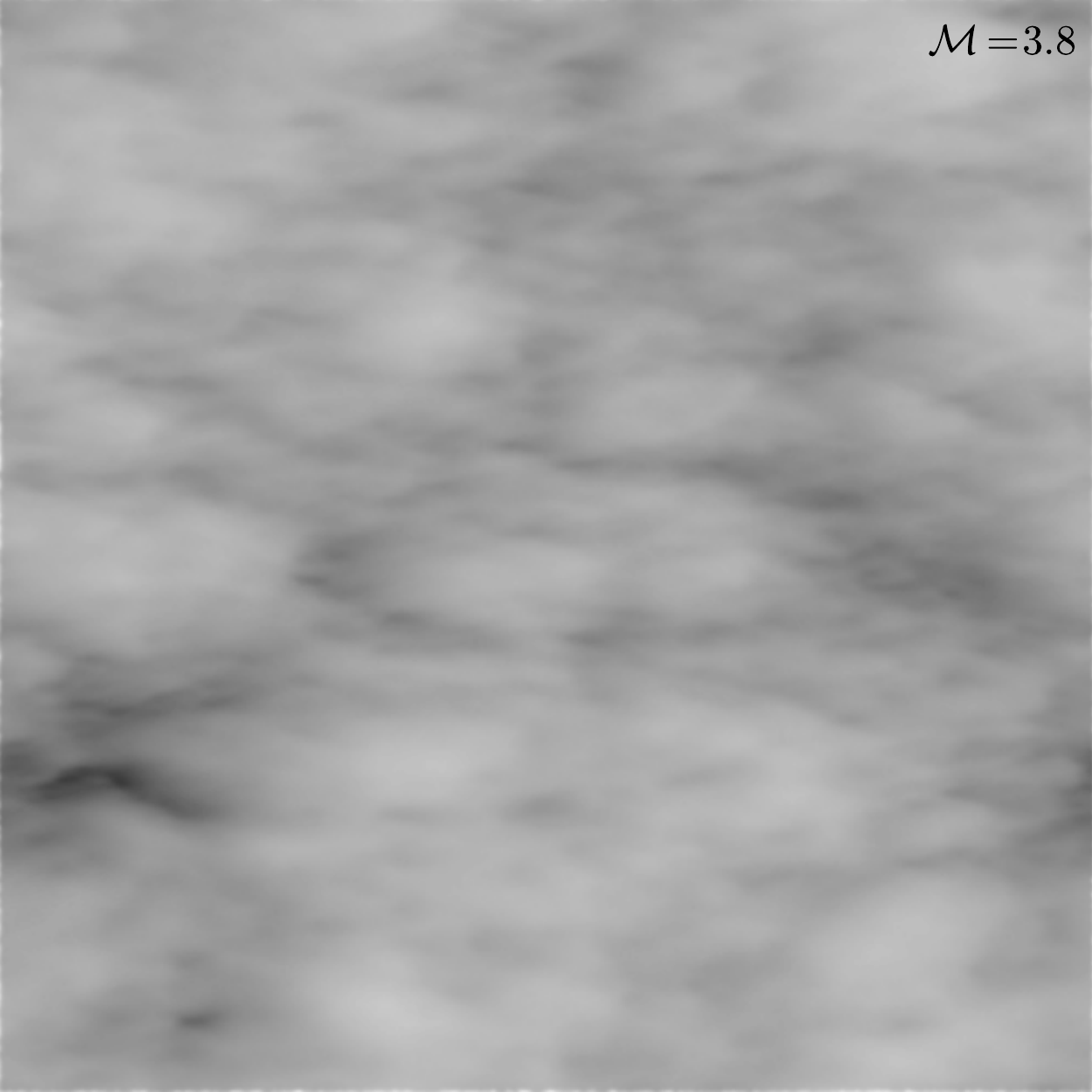}\\
\includegraphics[width=5cm]{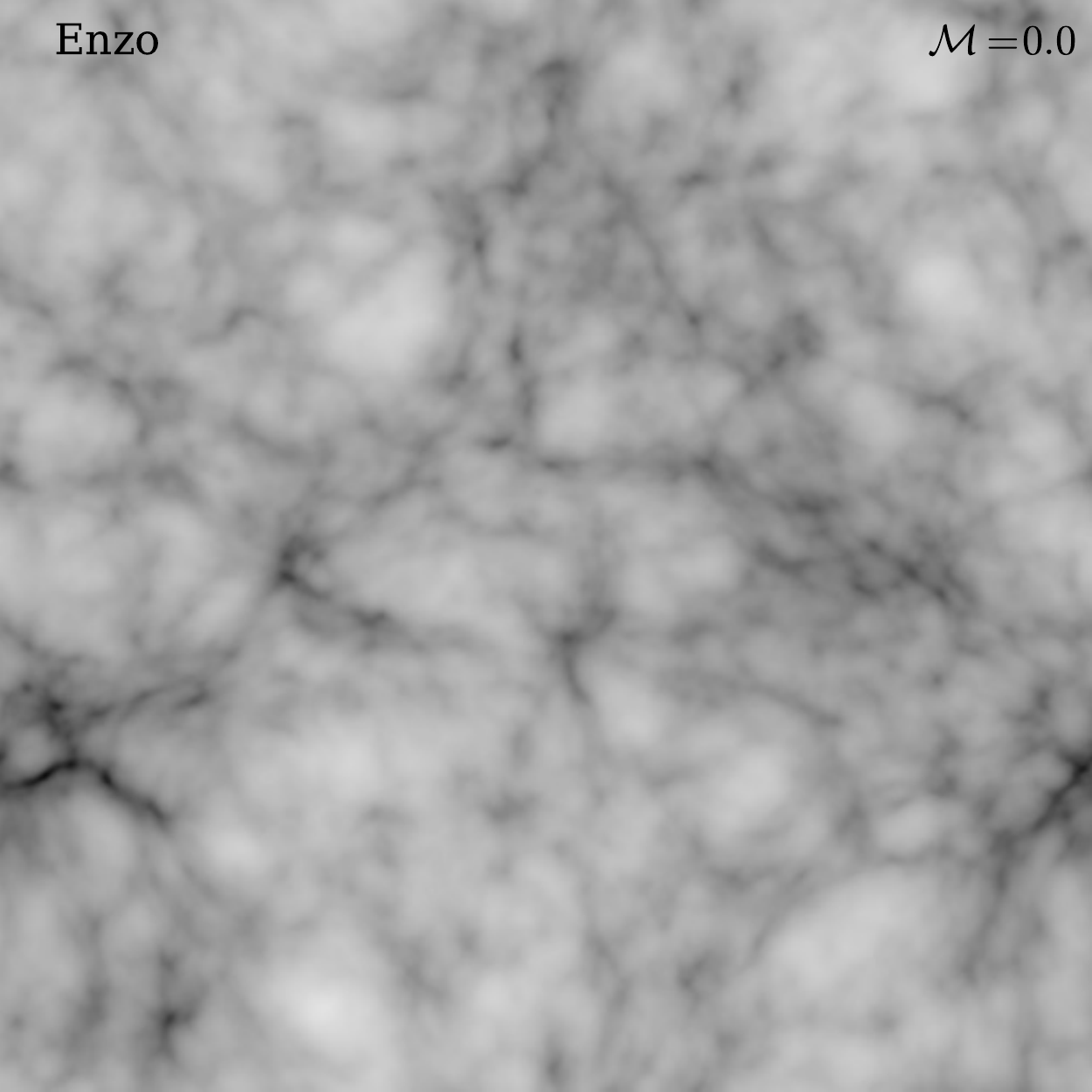}
\includegraphics[width=5cm]{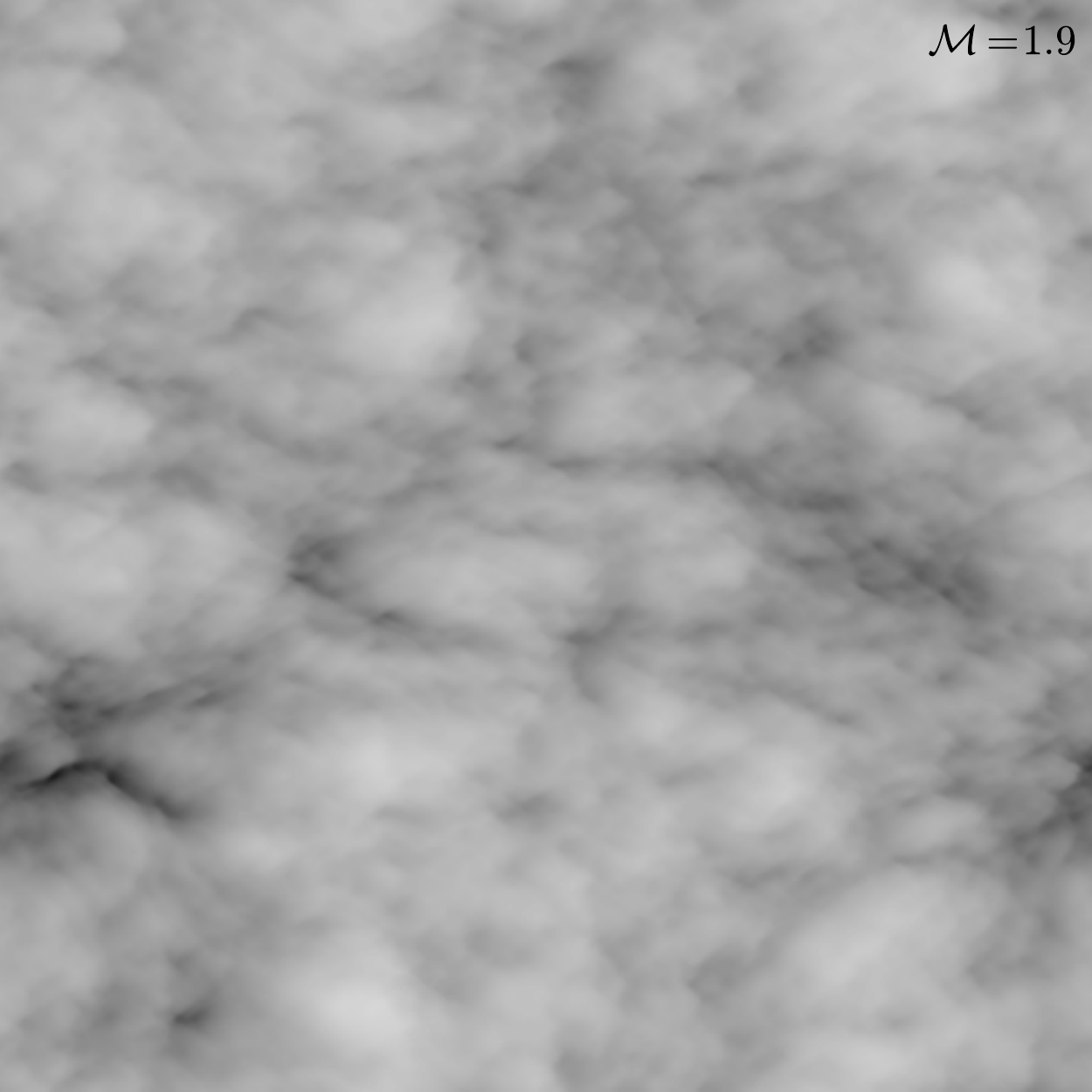}
\includegraphics[width=5cm]{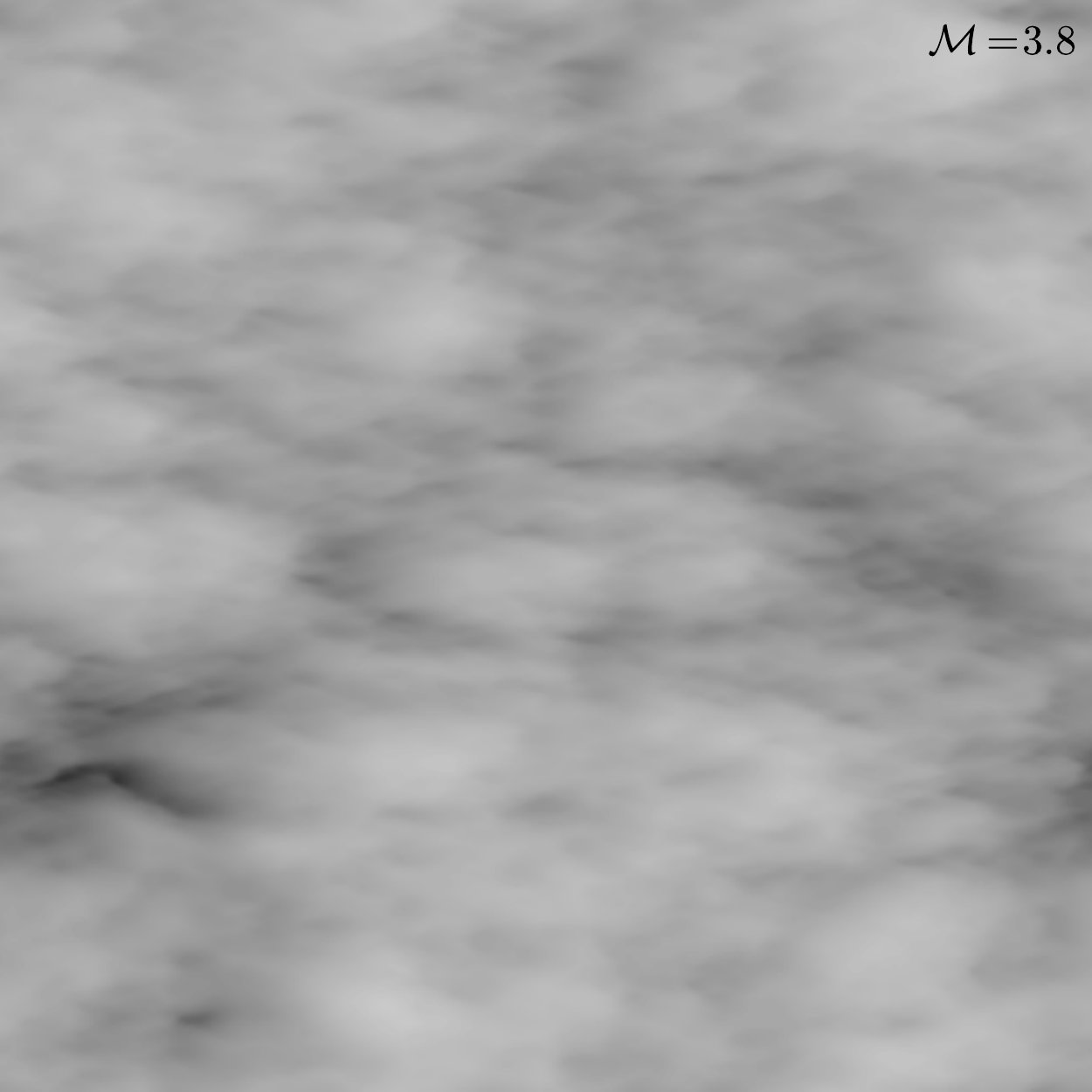}
\end{center}
\caption{Slices of $\log(1+\delta_b)$ through $z=20$ snapshots of the \{$0.2$~Mpc$/h$,~
  $2\times256^3$~resolution~element\} GADGET (top panels) and Enzo (bottom panels) simulations.  The top left, middle, and right panels correspond respectively to the cases of $v_\bc =0$ [$\calM_\bc = 0$], $v_\bc =30
  \,(z/1000)$~km~s$^{-1}$ [$\calM_\bc = 1.9$], and $v_\bc =60 \,
  (z/1000)$~km~s$^{-1}$ [$\calM_\bc = 3.8$]. 
 Dark regions represent overdensities and light underdensities, and
  the contrast is the same in all of the panels.  These simulations
  were initialized with the same random numbers and such that the
  baryons were flowing to the right in the cases with nonzero $v_\bc$.
  The Enzo simulations used a fixed uni-grid with $512^3$
  cells. \label{fig:images}}
\end{figure*}

The major improvements of our algorithm over other cosmological initial conditions generators are:
\begin{itemize}
\item The baryons and dark matter have distributions and velocities that are consistent with linear theory, including the impact of pressure.  Other commonly used initial conditions generators assume that the two components trace each other (the GADGET publicly available initial conditions code), that the dark matter and baryon velocities are proportional to each components' respective transfer function (the Enzo distribution ``inits'' code), or that the two components have the same velocity \citep{yoshida03}.   Other codes assume some variant of $\grad  \dot \phi = \Omega_m(z)^{5/9} H \, \grad \phi$, which is not valid at scales where pressure is important nor when radiation impacts cosmological expansion.
\item We include the effects of radiation in the initial densities and velocities.  Radiation is also included in the background evolution in the simulation themselves.  For both, we assume that the three species of neutrino are relativistic at all redshifts.  \change{ Radiation impacts the rate of growth at the 10\% level for simulations starting at $z\sim300$.}
\item The mean temperature and electron density are initialized with
  the values calculated with the RECFAST recombination code \citep{seager99}.\footnote{Compton cooling is
  also self-consistently included in the simulations.  GADGET and Enzo first stars
    calculations appear to use Case~A recombination coefficients, but
    Case~B (and calibrated for low temperatures) is a better choice
    and is required to obtain the correct evolution in the electron
    fraction and, hence, the correct thermal history.}  Our initial
  conditions also include fluctuations in the gas temperature as
  calculated from linear theory, \change{ an improvement that \citet{naoz07} stressed as important.}\footnote{The prefactor for converting
    temperature units in the Enzo code appears to have a typo (1.88e6
    should read 1.8182e6), resulting in an absolute error in the
    temperature of $\approx 3.4\%$. We have not corrected our
    calculations for this error.}
\item When a relative bulk velocity between the dark
  matter and baryons ($v_\bc > 0$) is required, we use linear solutions that self-consistently incorporate $v_\bc$.  This improvement is emphasized in the ensuing discussions.
\end{itemize}
 This initial conditions generator, the Cosmological Initial Conditions for AMR and SPH Simulations ({\sc CICsASS}) can be downloaded at \url{astro.berkeley.edu/$\sim$mmcquinn/codes}.

\subsection{Numerical Simulations}
\label{sec:nsims}

\begin{figure}
\begin{center}
\includegraphics[width=3in]{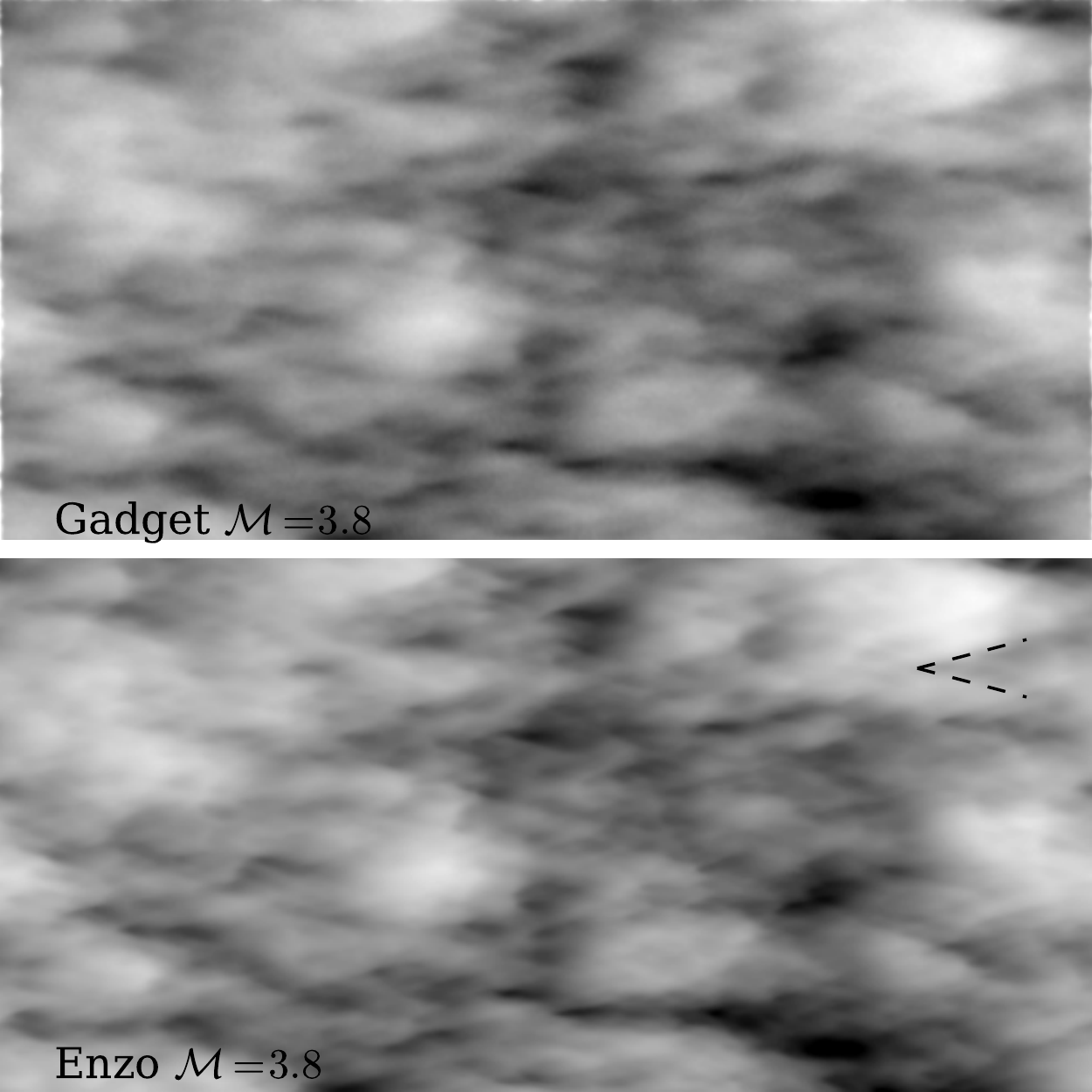}
\end{center}
\caption{Slices through the GADGET (top panel) and Enzo (bottom panel) simulations.  Shown is the same slice through
  $\log(1+\delta_b)$ in the left panels in
  Fig.~\ref{fig:images}  (the case with $\calM_\bc = 3.8$), except that only the top half of
  Fig.~\ref{fig:images} is shown and the contrast is linear for $0.5\lesssim
  \delta_b \lesssim 2.0$. The dashed, rotated `V' in the bottom panel shows the opening angle of a bow shock for $\calM_\bc = 3.8$.  \label{fig:Machcone}}
\end{figure}

We use these initial conditions with the cosmological codes GADGET3
\citep{springel01} and Enzo v2.1.1 \citep{oshea04} with the default
piecewise parabolic method to solve the equations of hydrodynamics.
GADGET solves the equations of fluid dynamics with the smooth particle
hydrodynamics (SPH) method, whereas Enzo is a grid code with adaptive
mesh refinement (AMR).  Furthermore, for gravity GADGET uses a the
tree--particle mesh grid whereas Enzo a nested particle mesh grid.
Both the GADGET and Enzo codes have been shown to conserve entropy at
the part in $1000$--level for the expansion of a homogeneous Universe
\citep{oshea05}.  Such accuracy is unusual among hydrodynamics codes,
and owes to the entropy--conserving formalism of GADGET and the
$3^{\rm rd}$--order accurate in space, $2^{\rm nd}$--order in time
Riemann solver employed by Enzo.  In addition, we require the codes to
capture the weak shocks that develop from the mildly supersonic flows
in the simulations.  Shock capturing is a strength of the Enzo
algorithm but a potential weakness of GADGET (and SPH codes in
general), which does not explicitly capture shocks.  Hence, our work
includes a direct comparison between these two codes.

The GADGET simulations that were run include \{box size in
Mpc$/h$,~gas particle number\} of $\{0.1,~128^3\}$, $\{0.1,~256^3\}$,
$\{0.2,~256^3\}$, $\{0.2,~512^3\}$, $\{0.5,~512^3\}$, and
$\{1,~768^3\}$, all initialized at a redshift of
$z_i=200$.\footnote{The $768^3$ simulations use the more sophisticated
  viscosity implementation of \citet{morris97}.}  \change{Our GADGET
  simulations were run using adaptive gravitational softening of the
  gas particles to reduce the amount of gas--dark matter particle coupling (Appendix \ref{sec:partcoup}).}  At each box size and particle number, baryon streaming velocities, $v_{\rm bc}$,
that result in Mach numbers of $\calM_\bc = 0$ and $1.9$ during
the Dark Ages were both seperately run.  (Note that, at the scales captured in our simulation
boxes, the baryons were initially moving coherently with velocity
$v_\bc(z_i)$ as a uniform wind.  The spatial distribution of $v_\bc$
is determined by the photon--baryon acoustic physics for which the spatial fluctuations are damped on scales captured by our simulation boxes by Silk Damping.) These choices allowed us to
explore the sensitivity of our results to resolution and sample
variance.  Note that all simulations resolve to varying degrees the
Jeans' scale (Fig.~\ref{fig:char_scales}), and the dark matter (gas)
particle mass in the $\{0.5,~512^3\}$ simulation is $82 ~\Msun$ ($17~
\Msun$).  We also reran a sample of these simulations with (1)
$\calM_\bc = 0.6$, (2) $\calM_\bc = 3.8$, (3) a more sophisticated
separate artificial viscosity implementation, \change{(4) with a fixed gravitational softening length for the gas, and (5)} $z_i=400$.
Finally, we do not adopt a standard practice in early Universe
simulations of increasing $\sigma_8$ to compensate for missing
large-scale power.  We found this artifice makes the results difficult
to interpret.

The Enzo simulations we ran include runs with a uniform grid
(with no AMR) with \{box size in Mpc$/h$,~grid size\} of
$\{0.1,~256^3\}$, $\{0.2,~256^3\}$, $\{0.2,~512^3\}$, and
$\{0.5,~512^3\}$ with both $\calM_\bc = 0$ and $1.9$ (and in a couple
cases $3.8$).  We also ran AMR runs with $\{0.2,~256^3\}$ and
$\{0.5,~512^3\}$ with $4$ levels of adaptivity for both the hydro
grid and for the gravity grid. We refined on gas density and dark
matter density when the mass in the cells exceeded 2$\times$
(4$\times$) the mean density in the $0.2~\cMpc/h$ ($0.5~\cMpc/h$)
simulation.  In all of the GADGET simulations and all of the adaptive runs of
the Enzo simulations, molecular hydrogen formation was not included.
Molecular hydrogen is the only active coolant at the gas temperatures in our simulations, and its absence prevents the runaway collapse into stars of dense gas in the first halos.

We have subjected our simulations to a battery of tests in order to confirm the robustness of our results.  For both GADGET and Enzo, we looked at the grid size, box size,
 maximum time-step size, number of
 particles/cells, and frame of reference for
 the relative velocity of the baryons and dark matter on the grid.  We found that the statistics we considered, such as the volume-averaged temperature and the power spectrum of the matter, were robust to these choices aside from grid size and box size (and we show these dependencies in ensuing plots).  For Enzo, we also investigated different Riemann solvers (and saw no differences) and the number of AMR levels.
 \change{For GADGET, we looked at the impact of using an adaptive versus a fixed gravitational softening length for the gas, as well as different ways of staggering the initial distribution of particles and different artificial viscosity implementations.\footnote{Our parametrization of the \citet{morris97} artificial viscosity results in a $10\times$ smaller artificial viscosity with $\alpha_v = 0.1$ in locations greater than $2$ smoothing lengths from where this parametrization triggers as having sufficiently larger $\grad \cdot \bfv$.  Our standard implementation assumes $\alpha_v = 1$ everywhere, using the default GADGET implementation described in \citealt{springel01}.}   We found that, when using an adaptive gravitational softening length, there was excellent convergence and agreement between GADGET and Enzo (see Paper II).  However, with a fixed gravitational softening length, coupling between the dark matter and gas particles spuriously modifies the temperature of the gas as discussed in Appendix B.
}

 We ran the simulations on a combination of computing resources
 including the Henyey computing cluster at University of California, Berkeley, and the XSEDE
 cluster Trestles. The $\{0.5,~512^3\}$ GADGET simulation required
  $6000\,$cpu-hr on 128 cores of the Henyey cluster to reach
 $z\approx 10$, whereas the $\{0.5,~512^3\}$ Enzo uni-grid simulation
  required $600\,$cpu-hr to reach $z\approx 10$.  The
 $\{0.5,~512^3\}$ Enzo simulation with four levels of AMR required $\gtrsim
 10^4\,$cpu-hr on 128 cores to reach $z=20$. 

\section{Results I: Initial Findings}
\label{ss:firstresults}

The top left, top middle, and top right panels in Figure~\ref{fig:images} show slices of $\log(1+\delta_b)$ through
three \{$0.2$~Mpc$/h$,~ $2\times256^3$~particle\} GADGET simulations
initialized with $v_\bc =0$, $v_\bc =30 \,(z/1000)$~km~s$^{-1}$ [or Mach number $\calM_\bc  = 1.9$], and
$v_\bc =60 \, (z/1000)$~km~s$^{-1}$  [$\calM_\bc = 3.8$], respectively.  Here, $\delta_b$ is the gas overdensity.  Note that 60\% of the cosmic volume would have had $\calM_\bc > 1.9$ and $6\%$ would have had $\calM_\bc > 3.8$ in the concordance $\Lambda$CDM cosmological model.  The most commonly simulated $\calM_\bc = 0$ case is the least representative: only $10\%$ of space had $\calM_\bc < 1$.  The bottom panel shows the same slices but through the Enzo
simulations.  Both the GADGET and Enzo simulations were initialized with the same random
numbers and such that the baryons were flowing to the right for the
cases with $v_\bc>0$.  Note that while the large-scale structures
largely line-up, there are marked differences between the three cases,
with larger $v_\bc$ leading to a damping of the filamentary structures
perpendicular to the flow direction.   The supersonic flows also lead to the development
of Mach cones around the dark matter
substructure.  Figure \ref{fig:Machcone} illustrates the prominence of these cones, showing slices of density in the $\calM_{\rm bc}=3.8$ simulations.  The Mach cones are most striking in movies that pan through slices in the box. Paper II investigates whether these shocks heat the IGM.

In the remainder of this section, we compare the the evolution of the matter in the GADGET and Enzo
simulations both at times where the evolution is linear (such that they can also be tested against analytic solutions) and also at more advanced stages.

\subsection{Linear Evolution in Simulations}
\label{ss:lin}

\begin{figure}
\includegraphics[height=\columnwidth]{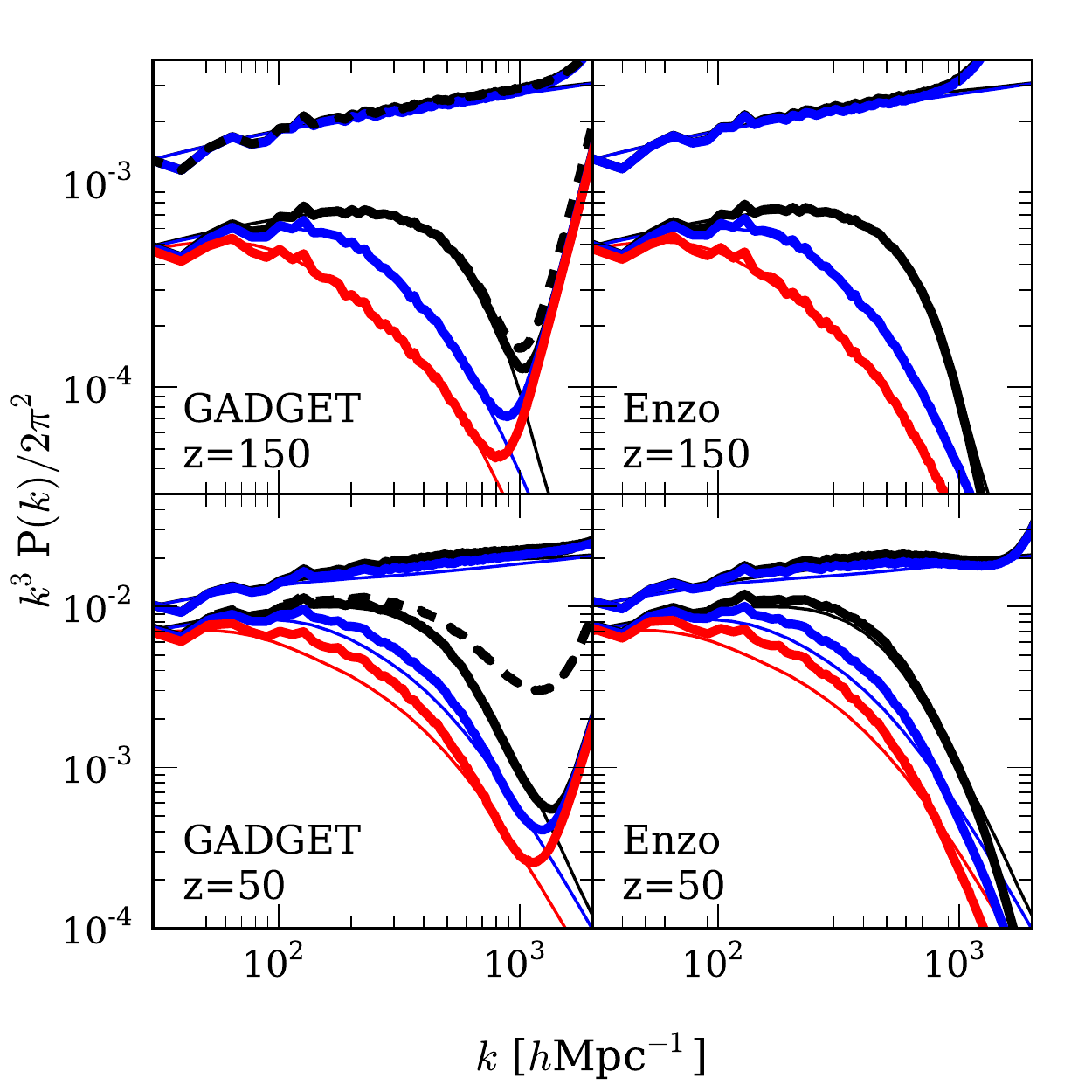}
\caption{Simulated dark matter (top curves) and baryonic power spectra (bottom curves) for the cases $\calM_{\rm bc} = 0$, $\calM_{\rm bc} = 1.9$, and $\calM_{\rm bc} = 3.8$ (black, blue, and red thick solid curves, respectively). The thin solid curves with the corresponding color are the linear theory solutions for these $\calM_{\rm bc}$.  The curves are calculated from the $0.5~\cMpc/h$, $512^3$ dark matter particle GADGET and Enzo simulations (left and right panel, respectively).  The turn-up in power at $k>10^3~\cMpc^{-1}$ owes to shot noise. 
 \label{fig:ltcomp}}
\end{figure}

The most basic test for whether the simulations are behaving properly
is whether they are capturing linear theory at times when it should
hold.  Linear theory is a good approximation for the matter power
spectrum in the simulations from initialization ($z=200$ unless stated
otherwise) until $z = 50$, at which point the dark matter and baryonic
power spectra deviate at the order unity--level from its predictions.
Figure \ref{fig:ltcomp} shows this comparison for different
initialization methods and codes, comparing the power spectrum in the
GADGET (left panel) and Enzo with AMR (right panel) $0.5~\cMpc/h$,
$2\times512^3$~base-grid resolution element simulations at $z=150$ and
$50$.  We find that AMR with our refinement criteria makes little
difference at these redshifts. The thick curves in Figure
\ref{fig:ltcomp} are the dark matter and baryonic power spectra from
these runs, and the thin curves are the predictions of linear
theory. To calculate these curves, the particles in each snapshot
were placed onto a Cartesian grid with cloud-in-cell interpolation.
We then divided out the cloud-in-cell window function when calculating
the power spectrum.  The gas density from Enzo was extracted onto a
fixed grid at its coarsest resolution. We then divided out the
nearest-grid-point window function when calculating each power
spectrum.

At $z=150$, both
codes trace the linear theory predictions well, with both the baryon and
dark matter components in GADGET and just the dark matter in Enzo
deviating at the highest wavenumbers owing to shot noise.
 Whereas, the gas in Enzo underpredicts linear theory at wavenumbers that correspond to a few times smaller than the Nyquist of the root grid. 
  However, by $z=50$ in both GADGET and Enzo, there are notable deviations from linear theory, \change{especially when $\calMbc > 0$}, with the dark matter power spectra in Enzo noticeably smaller than GADGET at the smallest wavenumbers.

 \change{The thick dashed curve represent the power spectrum of the
  baryons in GADGET for $\calM_\bc = 0$ when adopting the standard practice of running with a fixed gravitational softening
  length.  Much of this deviation at $z=50$ between the $\calM_\bc = 0$ linear theory curve and this curve is spurious and owes to gas particle--dark matter particle coupling.}
Because the gas overdensity fluctuations are just $10^{-2}$ at $z=50$, even a small amount of numerical coupling
between the dark matter and baryonic particles can source significant
fluctuations. 
Appendix B provides additional discussion of particle coupling and its impact in the simulations.  The amount of coupling does not depend on the box size or particle number of the simulation, such that this coupling sneakily evades simple convergence tests and can be easily confused with nonlinear evolution. \change{This particle coupling is eliminated by using adaptive gravitational softening lengths for the baryons.} 
\change{This coupling does not only impact simulations of the $z\sim 20$ universe, and, for example, will impact SPH simulations of the Ly$\alpha$ forest.}

\change{With adaptive softening in GADGET}, the deviations from linear theory between the GADGET and Enzo power spectrum curves largely agree at $z=50$.  This suggests that the simulations in both codes are capturing linear theory over the realm where it applies and also that these deviations are real and linear theory is starting to error by $z=50$, \change{especially for $\calMbc > 0$.}  In addition, the matter power spectra in the simulations we ran with other box sizes, initialization redshifts, and artificial viscosity implementations agree well with the simulations featured in Figure \ref{fig:ltcomp}, both at $z=50$ and at $z=150$. %

\subsection{Nonlinear Evolution}
\label{ss:nonlin}
\begin{figure}
\rotatebox{-90}{
\includegraphics[height=\columnwidth]{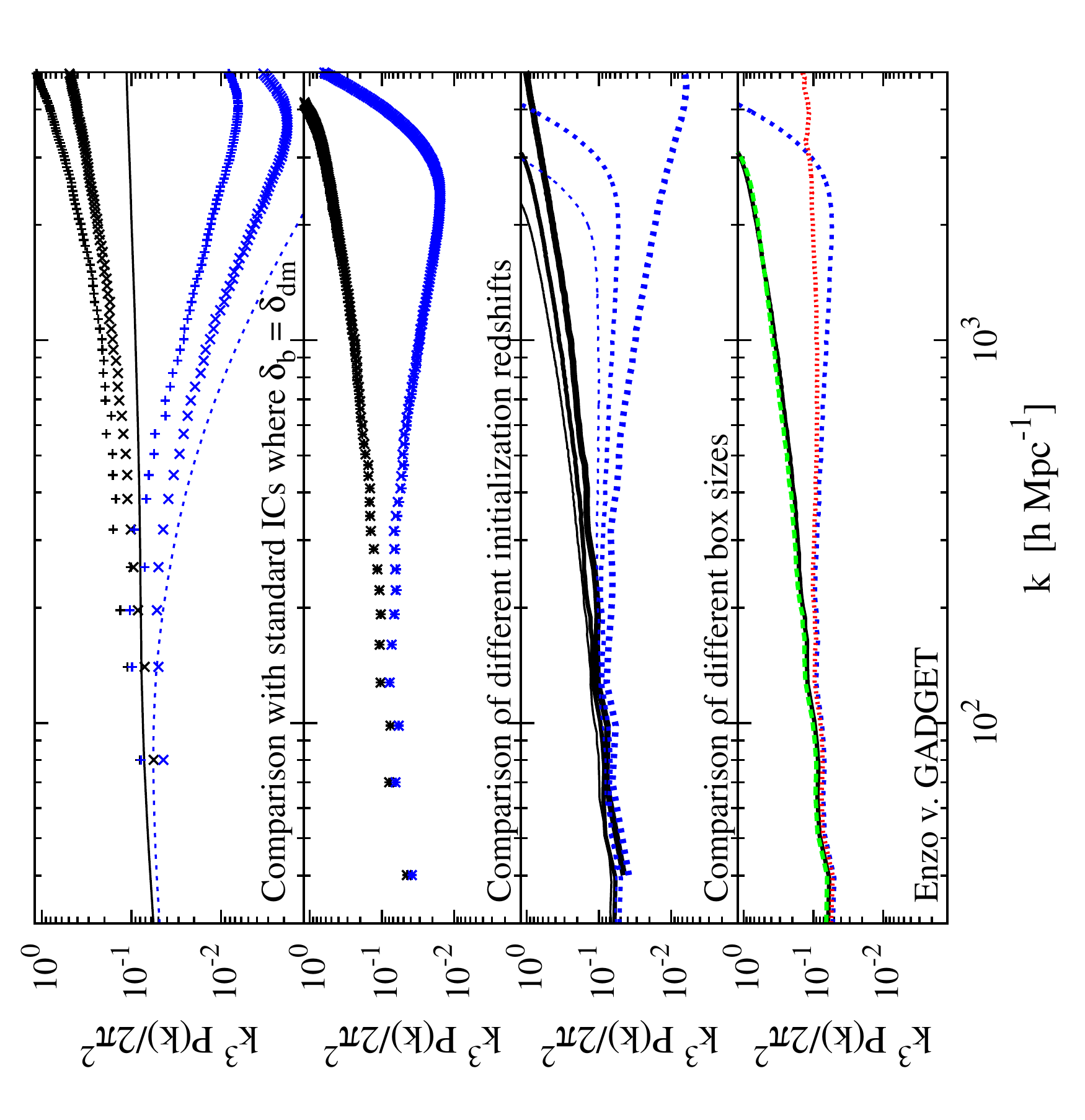}}
\caption{Density power spectra from different simulations at $z=20$, all with $\calM_{\rm bc} = 1.9$ and using the GADGET code \change{with a fixed smoothing length}  and initialized at $z=200$, unless stated otherwise.  Top panel:  Comparison of our initial conditions generator (crosses) with initial conditions using the common practice of setting $\delta_b$ and $\delta_{c}$ to the total matter over density (pluses).  Black signifies the dark matter power and blue the baryonic power, and the curves represent the linear-theory predictions.  Second panel: The same but instead comparing the simulation with our fiducial initial conditions (pluses) with one initialized at $z=400$ (crosses).  Note that the two cases largely overlap.  Third Panel:  Comparison of the impact of box size.  Shown are our largest particle number simulations that are run with box sizes of $\{0.2,~0.5,~1.0\}~\cMpc/h$ in order of decreasing line width.  The dashed curves are the baryonic power, and the solid curves are the dark matter power.  Bottom Panel:  Comparison of gas and dark matter power spectra between the GADGET and Enzo simulations with AMR, both with specifications of $0.5~\cMpc/h$ and $512^3~$initial gaseous resolution elements.  The dashed green curve in the bottom panel is the dark matter power spectrum from the Enzo simulation, and the dotted red curve is the same but instead the baryonic power spectrum. \label{fig:ltcomp2}}
\end{figure}

Ultimately, we want to study the nonlinear behavior of the gas and dark matter in the cosmological simulations: e.g., shocking, collapse of structure, and the formation of the
 first stars.  An important step towards this comparison is to verify that the nonlinear solutions of our simulations agree.  This amounts to comparing the density field of the simulations at $z\lesssim 50$, times when some baryonic density fluctuations begin to go nonlinear and when the first gas-rich halos collapse.  Figure \ref{fig:ltcomp2} shows the dark matter (black markers) and
baryonic (blue markers) power spectra at $z=20$.  The top panel compares a
simulation with our initial conditions generator (crosses) to
one that uses initial conditions with $\delta_b$ and $\delta_{c}$ both set equal to the total matter overdensity
(pluses), where $\delta_b$ and $\delta_c$ are the overdensities in gas and dark matter, respectively.  In addition, the curves in the top panel
show linear theory, which only captures the growth of the larger modes in these simulations.  This panel shows that the common practice of starting a simulations of the first stars with $\delta_b = \delta_{c}$ results in an
overshoot in the size of density fluctuations with a fractional error of
${\cal O} (1)$.  This comparison stresses the importance of using different transfer functions for the two components, as also emphasized in \citet{naoz11}.  At
higher redshifts than are shown, we find that the power spectrum of the case initialized with $\delta_b = \delta_{c}$ contains
significant acoustic oscillations, an artifact from these over-pressurized initial conditions.

The second and third panels down in Figure \ref{fig:ltcomp2} compare the convergence between different initialization redshifts and box sizes, respectively.
The second panel 
compares a GADGET simulation with our fiducial initial conditions initialized at $z=200$ (pluses)
with one initialized at $z=400$ (crosses).  Note that the two cases are in excellent agreement.
However, convergence is not as successful in box size as in initialization redshift.  The third panel in Figure \ref{fig:ltcomp2} compares the power spectra between GADGET simulations 
with box sizes of $\{0.2,~0.5,~1.0~\}\cMpc/h$, in order of decreasing linewidth.  The dashed curves are the baryonic power and the solid curves are the dark matter power.  The simulations do not demonstrate significant convergence with increasing box size.  \change{We suspect based on calculations discussed in Section \ref{sec:scales} that if we could run a larger box simulation that resolved the Jeans' length, we would conclude that the power spectrum in the $1.0~\cMpc/h$ is converged to $\sim 10\%$. }

Lastly, the bottom panel  in Figure \ref{fig:ltcomp} compares the gas and dark matter power between the GADGET and Enzo
simulations with box sizes of $0.5~\cMpc/h$ and $512^3$~initial gaseous resolution elements.  The
dark matter power spectra largely agree between these two simulations.  We suspect the small excess in the
baryonic power in Enzo at $k\sim 10^3~\cMpc^{-1}$  is spurious
and owes to power induced by the Enzo AMR algorithm.  To test this hypothesis, we
compared our simulations in smaller boxes (with a box size of
$100/h\,\ckpc$) with and without AMR.  We
found that for the case without AMR, the power spectra were a much better match to the
linear theory power spectra as the grid size was increased (for $z\gtrsim 50$).  With AMR, we found that spurious extra power induced at the scale of the AMR grid migrated to larger scales (smaller $k$) as time proceeded.

We have focused on the power spectrum as a diagnostic of the nonlinear evolution, but it is by no means the only measure of structure formation nor the most interesting one.  Subsequently, we will focus on how the first structures form in the simulations.  While this analysis will again provide a comparison of the codes, our focus will emphasize new physical insights.  

The most significant improvement of our simulations is that the gas is initialized with displacements and velocities that are generated from linear solutions that include $v_\bc$.  This improvement allows us to properly
follow the infall of gas onto the first cosmic structures.  The ensuing discussion investigates the impact of $v_\bc$ on the density distribution of IGM and on the first halos that form with deep enough potential wells to retain their gas and, in some cases, form stars.  The ensuing discussion has implications for the thermal evolution of the IGM, the formation of the first stars, and the $z\sim 20$ 21cm signal.\\

\section{Results II: Dynamical Friction in the Early Universe}
\label{ss:df}

The situation in the early Universe -- dark matter clumps moving through a more homogeneous gas with velocity $v_\bc$ -- is similar to the toy case used to derive Chandrasekhar's classic dynamical friction formula of a massive particle moving through a homogeneous sea of collisionless particles \citep{1943ApJ....97..255C,binney87}.  A similar formula has been shown to apply to the case of a particle moving through gas (e.g., \citealt{ostriker99} and \citealt{furlanettoshocks}).   The dynamical friction timescale for a halo of mass $m_h$ to lose its energy to the background baryons with density $\rho_b$ and Mach number $\calM_\bc$ is 
\begin{eqnarray}
t_{\rm df} &=& \frac{v_{\bc}^3}{4 \pi \log \Lambda \,G_N^2 m_h \rho_b},\\
                    &=& 50 {\rm ~Myr} \left( \frac{\calM_\bc}{1.8}\right)^3 \, \left(\frac{ \log \Lambda}{3} \frac{m_h}{10^5 ~\Msun} \right)^{-1},
                    \label{eqn:tdyn}
\end{eqnarray}
where $\log \Lambda$ is the so called `Coulomb logarithm'.  In the
gaseous case and for linear perturbations, there exists an analytic
solution for $\log \Lambda$ assuming a point-like perturber moving for
a finite time \citep{ostriker99}:
\begin{eqnarray}
\log \Lambda &=&  \log \left[ \frac{\calM_\bc \,R_{J}}{r_{\rm min}}\right] + \frac{1}{2} \log \left[ 1- \calM_\bc^{-2} \right], \label{eqn:logL}\\
 &\approx& 2.9 - \frac{1}{3}\log \frac{m_h}{10^5 ~\Msun} +  \log \frac{ {\lambda}^{-1} \, \calM_\bc^{0.4}\, \sqrt{1- \calM_\bc^{-2}}}{2^{-0.4}\sqrt{1- 2^{-2}}} \nonumber \\
 &&+~  \frac{1}{2} \log  Z_{20}, \label{eqn:logL2}
\end{eqnarray}
where $R_J$ is the Jeans' scale (or the distance a sound wave travels in the age of the Universe for $\delta = 0$, which replaces $c_s \, t$ in the original expression in \citealt{ostriker99}).  To reach equation (\ref{eqn:logL2}), we replaced $r_{\rm min}$ \change{(the minimum radius from which the drag force was included in the \citealt{ostriker99} calculation) with $0.35 \, \calM^{0.6}$ times the characteristic scale of an extended object.  This approximation was suggested in \citet{kim09} to generalize the \citet{ostriker99} formula to extended objects sourcing nonlinear density perturbations.}  For definiteness, we set this characteristic scale to be $\lambda \, r_{\rm vir}$, where $r_{\rm vir}(m_h)$ is the virial radius of a halo of mass $m_h$.   Note that with equations (\ref{eqn:tdyn}) and (\ref{eqn:logL}), $t_{\rm df}$ is at a minimum for $\calM_\bc = 1-2$ \citep{ostriker99}, which coincides with the most probable values for $\calM_\bc$ in the concordance cosmological model.

A Hubble time equals $280 \, Z_{20}^{-3/2}~$Myr at times when $\Omega_m \approx 1$.  Thus, equation (\ref{eqn:tdyn}) implies that the surrounding gas will be decelerated and fall into halos with $m_h\gtrsim 10^{4-5}~\Msun$ by $z\sim 20$.  This bound coincides roughly with the mass bound on halos that can overcome pressure and maintain their gas \citep{naoz09}. 

Significant dynamical friction occurs when nonlinear structure forms such that gravity becomes strong enough to generate ${\cal O}(1)$ perturbations in the gas on smaller scales than $R_J$, the distance a sound wave travels in a Hubble time.  
The fact that $\log \Lambda \sim 1$ reflects that there is not a large range of scales between the sizes of objects and how far sound waves can travel.   

\begin{figure}
\begin{center}
\rotatebox{-90}{
\includegraphics[height=\columnwidth]{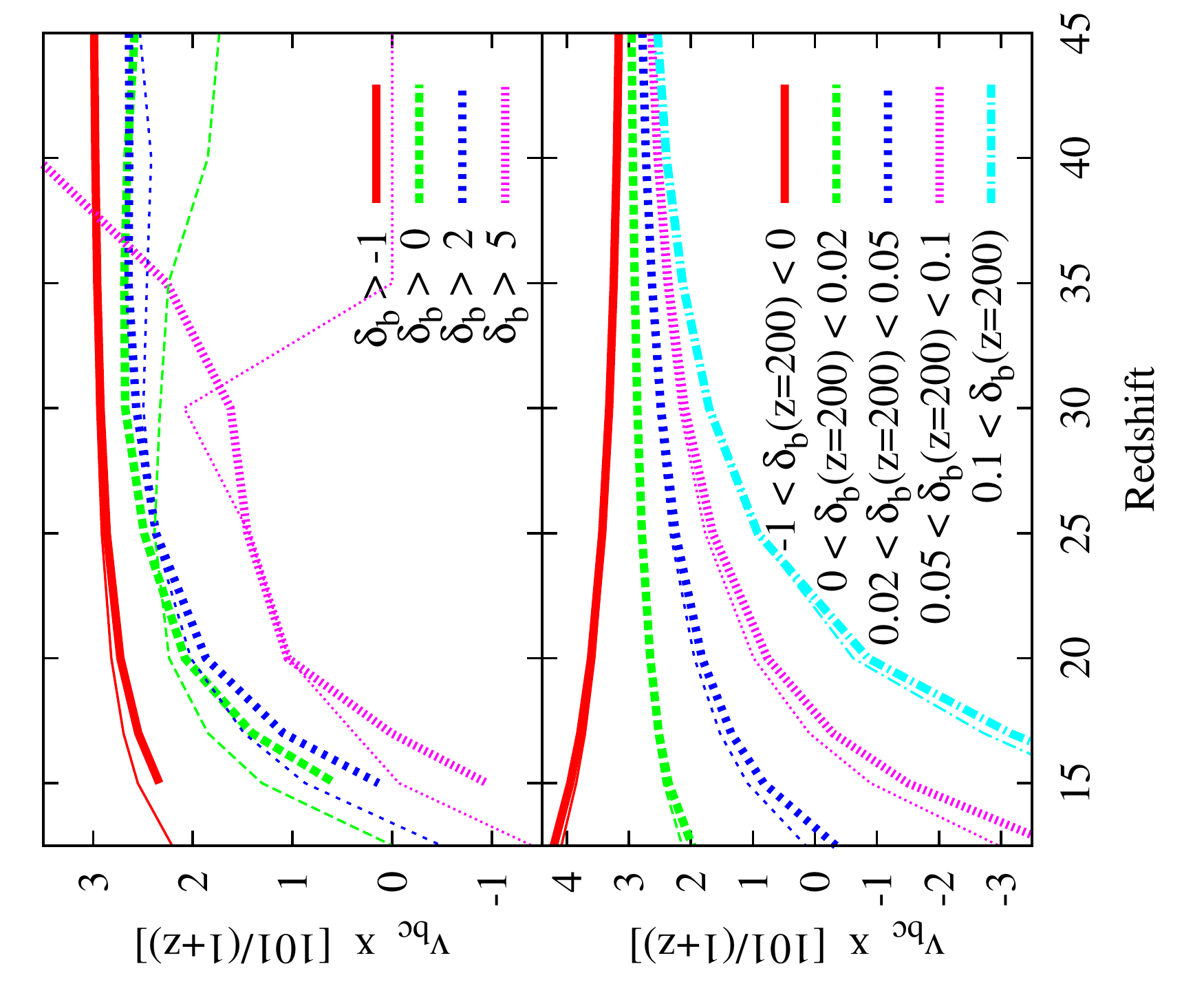}}
\end{center}
\caption{ Evolution of the component of the baryonic velocity in the direction of the baryonic flow times $101/(1+z)$.  Shown are different overdensity cuts in the $0.5~\cMpc/h$, $2\times 512^3$ (thick curves) and $0.2~\cMpc/h$, $2\times 512^3$ (thin curves) GADGET simulations with $\calM_\bc = 1.9$ ($v_\bc = 3~$km~s$^{-1}$ at $z=100$).  The top panel shows the average velocity above the specified overdensity thresholds at the specified redshift.  The bottom panel follows the velocity of a group of gas particles that fall with in the same density range at the time of initialization.  There are $52,~15,~21,~ 9,$ and $3\%$ of the particles in each density grouping specified in the bottom panel for the $0.5~\cMpc/h$ box.   Both panels illustrate how dynamical friction causes the gas and dark matter to decelerate into the same frame in overdense regions. \label{fig:vep}}
\end{figure}

Gaseous dynamical friction generates Mach cones around supersonic objects, in our case predominantly dark matter halos.  The shocks at the edges of these cones may heat the IGM as the halo is decelerated.  These cones are very visually apparent in the large-scale simulations with finite $v_\bc$ (see Fig.~\ref{fig:Machcone}) and especially when one zooms in on halos (as we will show in \S \ref{sec:results2}).  Figure \ref{fig:vep} plots the evolution of the average velocity of gas particles above a fixed overdensity threshold (top panel) and the average velocity of particles that fall into the same density range at initialization (bottom panel) in GADGET simulations with $\calM_\bc = 1.9$.  Both panels show that dynamical friction effectively acts to decelerate the densest regions.  Interestingly, the velocity even goes negative for the rarest, most overdense regions in the bottom panel, as the wake of baryonic material around a dark matter potential halts and falls back onto the dark matter potential well.  We find very similar trends in the simulations with $\calM_\bc = 0.6$ and $3.8$ as in the $\calM_\bc = 1.9$ simulations. 

\change{This dynamical fictional also causes the velocity of the dark matter halos to move in the direction opposite the decelerating baryons.  However, because there is $\approx 8$~times more mass in the dark matter, the change in velocity of dark matter overdensities from this dynamical friction is $\ll v_{\rm bc}$.}

\section{Results III: The Impact of $v_\bc$ on the First Structures and Halos}
\label{sec:results2}

\begin{figure*}
\begin{center}
\includegraphics[width=\textwidth]{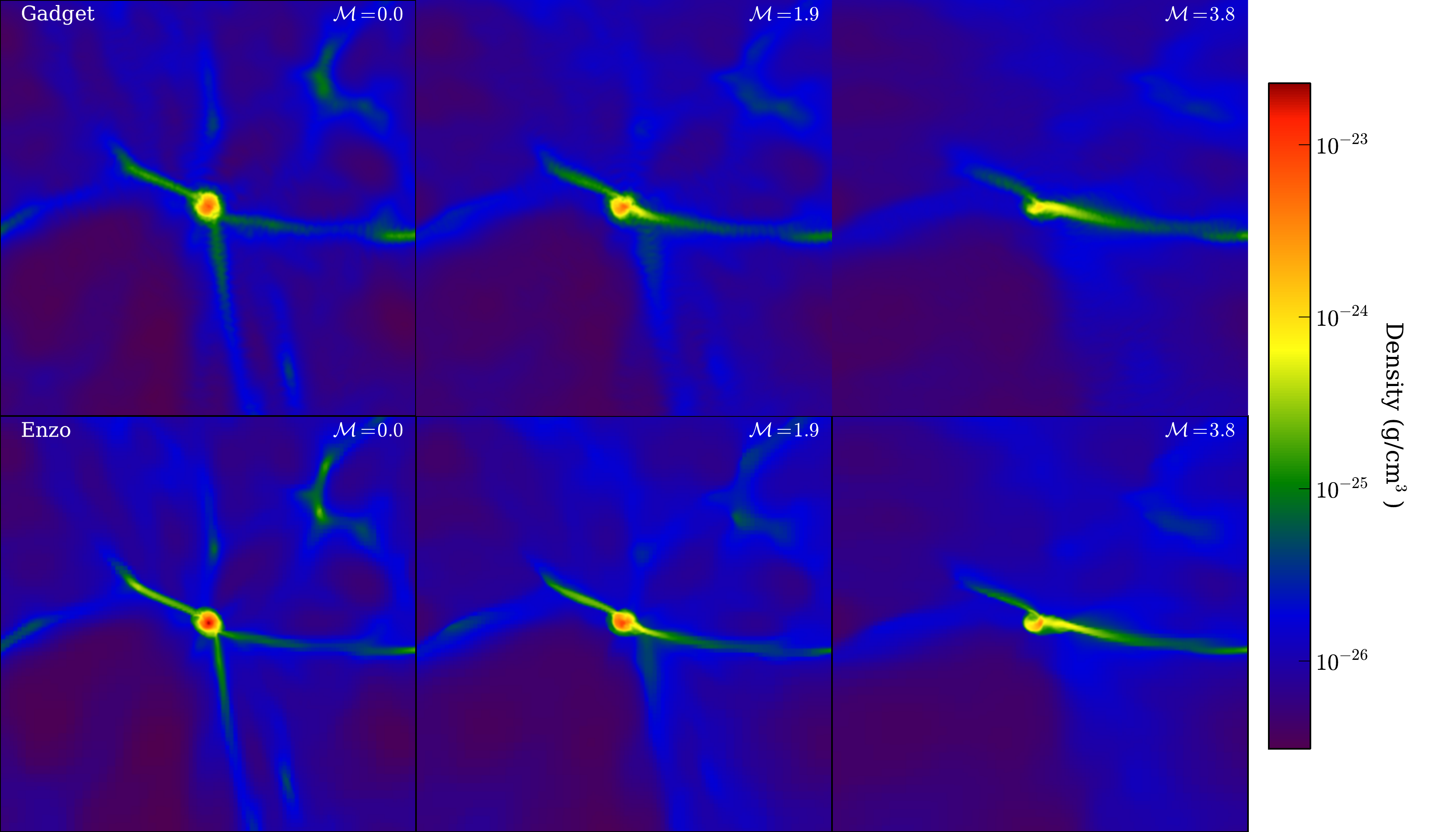}\\
\includegraphics[width=\textwidth]{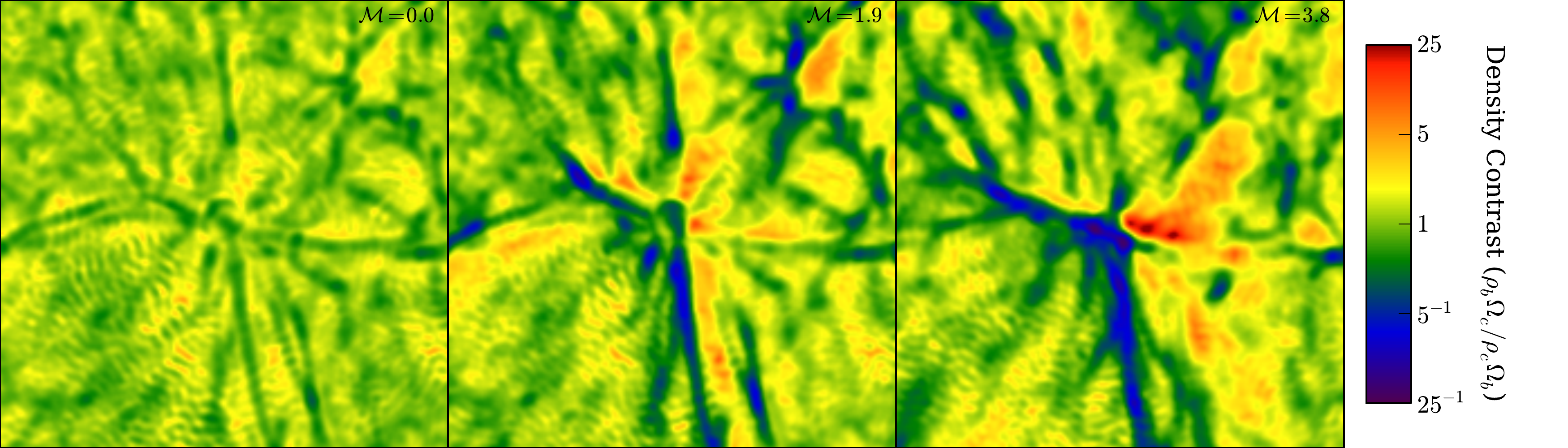}
\end{center}
\caption{Comparison of the gas density in the \emph{most} massive halo in the $\{512^3,~0.5~\cMpc/h\}$ simulations at
  $z=20$.  Plotted in the top (middle) row is a cut through this halo with width 50~$\ckpc/h$ in the GADGET (Enzo) simulation.    From left to right, $\calMbc = 0, 1.9,$ and $3.8$, with the baryons moving to the right.  This halo's friends--of--friends group mass in the $\calMbc = 0$ GADGET simulation is $2\times 10^6~\Msun$.    Plotted in the bottom row is the density
  contrast of the baryons to the dark matter or $(1+\delta_b)/(1+\delta_c)$ in the GADGET simulation.  Each slice is centered on the densest point within the halo.
 \label{fig:halo1} }
\end{figure*}

\begin{figure*}
\begin{center}
\includegraphics[width=\textwidth]{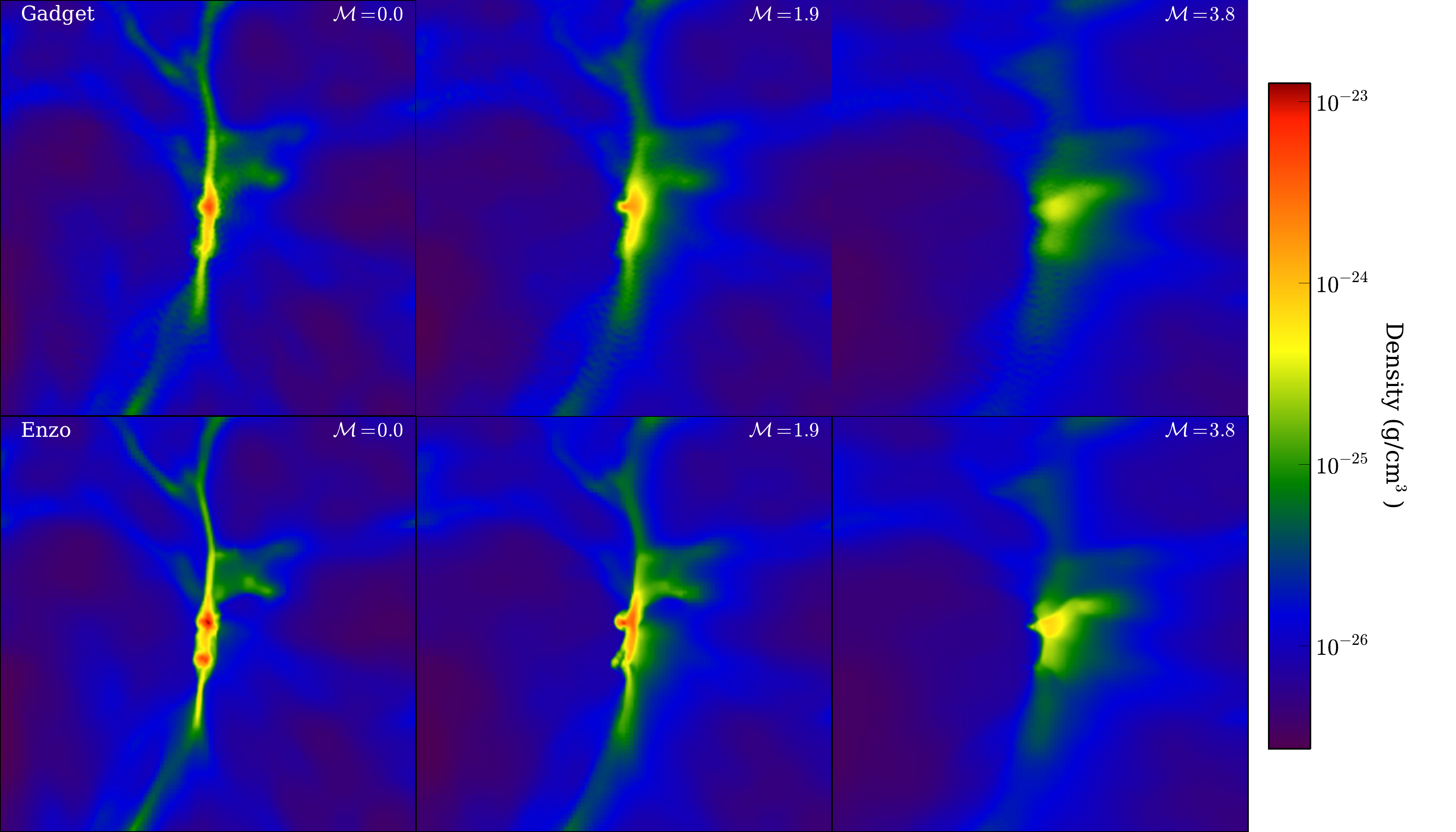}\\
\includegraphics[width=\textwidth]{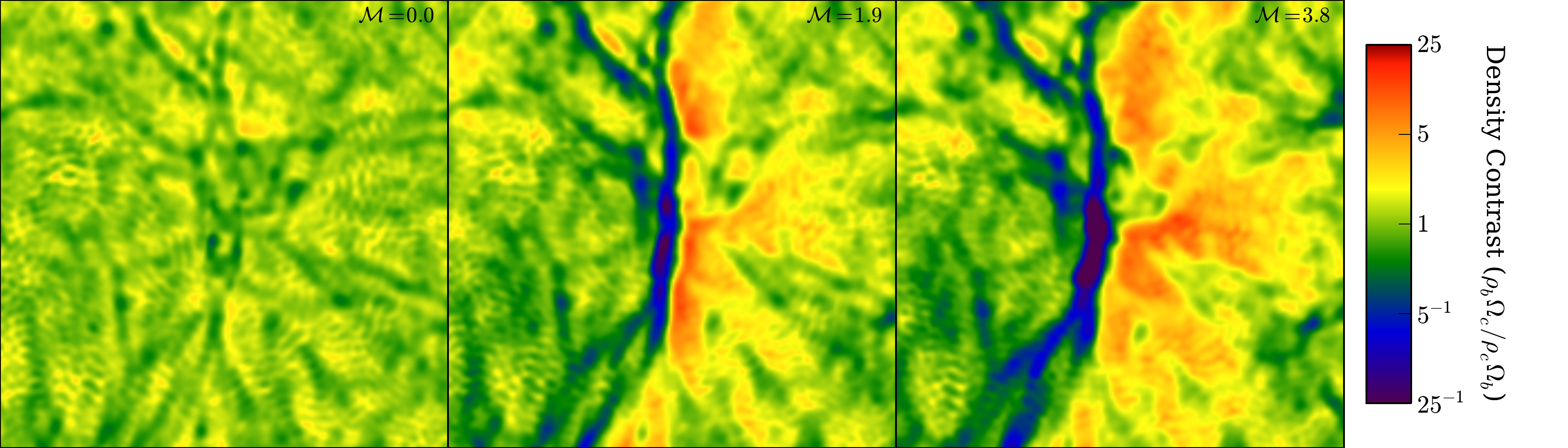}
\end{center}
\caption{Same as Fig.~\ref{fig:halo1}, but shown is the second most massive halo at $z=20$.  Its friends--of--friends group mass in the $\calMbc = 0$ GADGET simulation is $8\times10^5~\Msun$.\label{fig:halo2}}
\end{figure*}

\begin{figure*}
\begin{center}
\includegraphics[width=\textwidth]{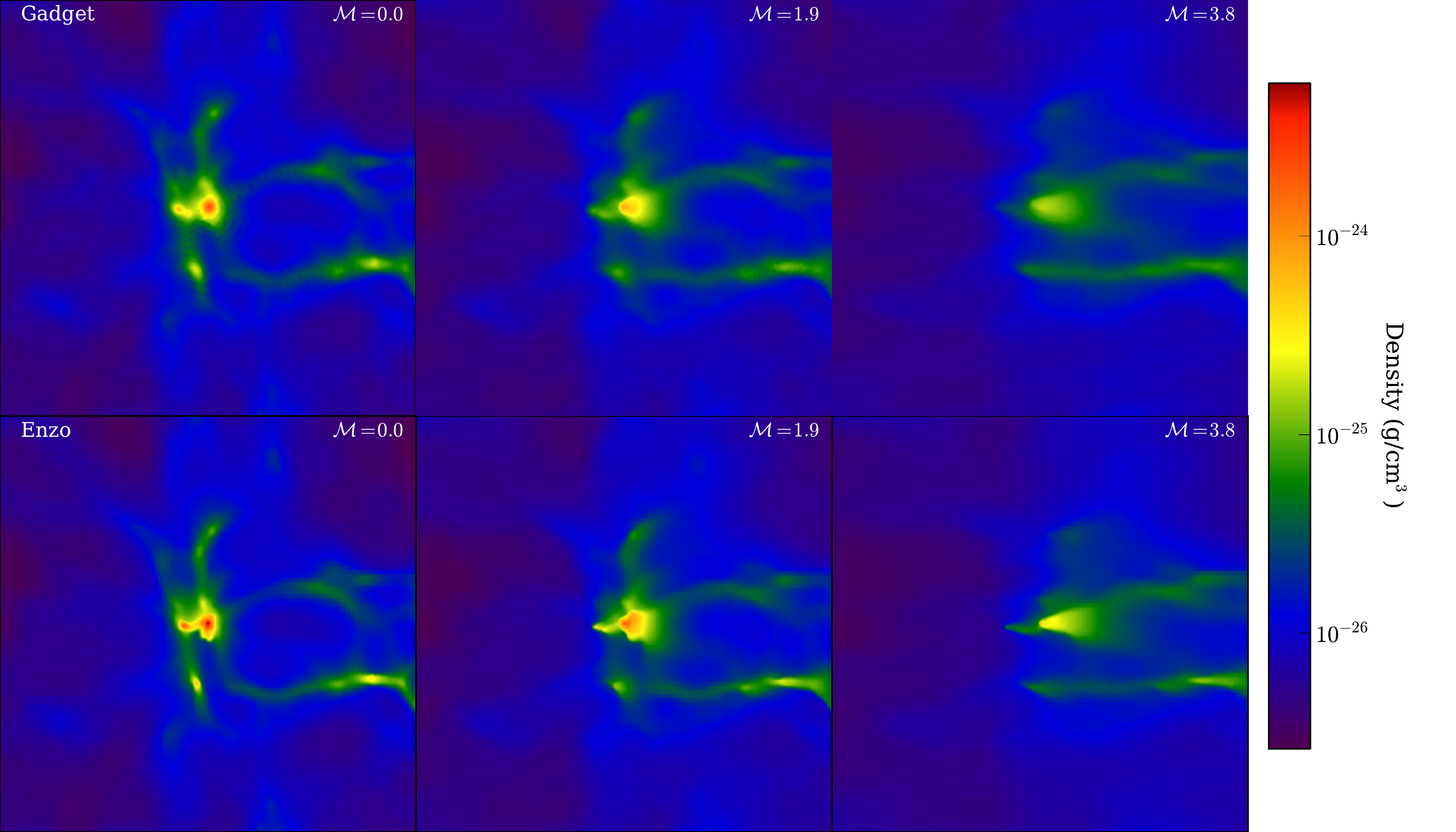}\\
\includegraphics[width=\textwidth]{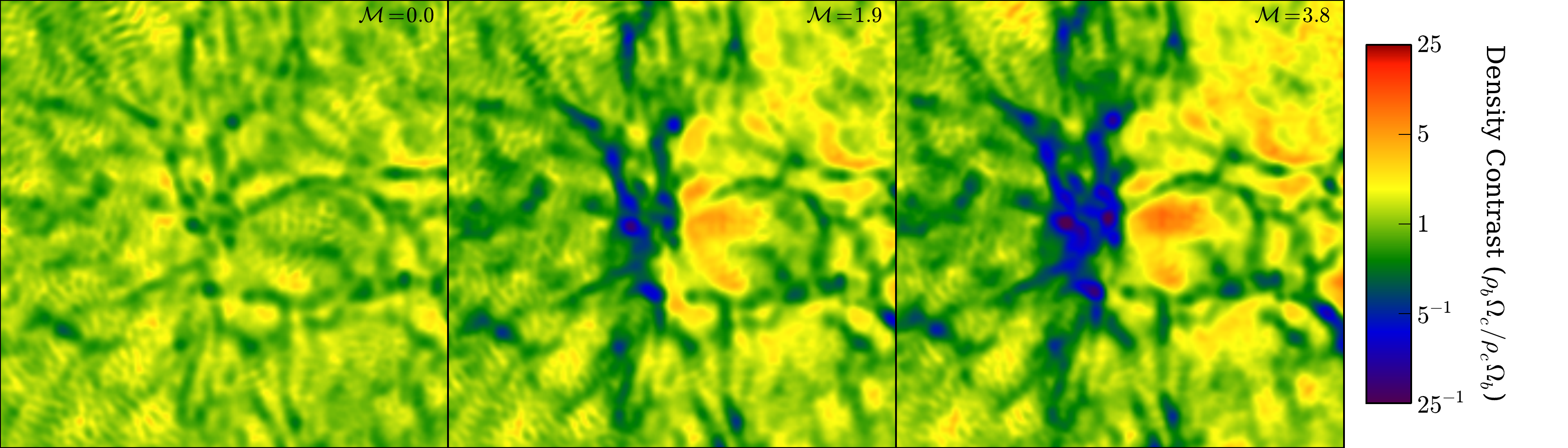}
\end{center}
\caption{Same as Fig.~\ref{fig:halo1}, but shown is the $\sim 20^{\rm th}$ most massive halo at $z=20$. Its friends--of--friends group mass in the $\calMbc = 0$ GADGET simulation is $2\times10^5~\Msun$.\label{fig:halo3}}
\end{figure*}

\begin{figure*}
\begin{center}
\includegraphics[width=\textwidth]{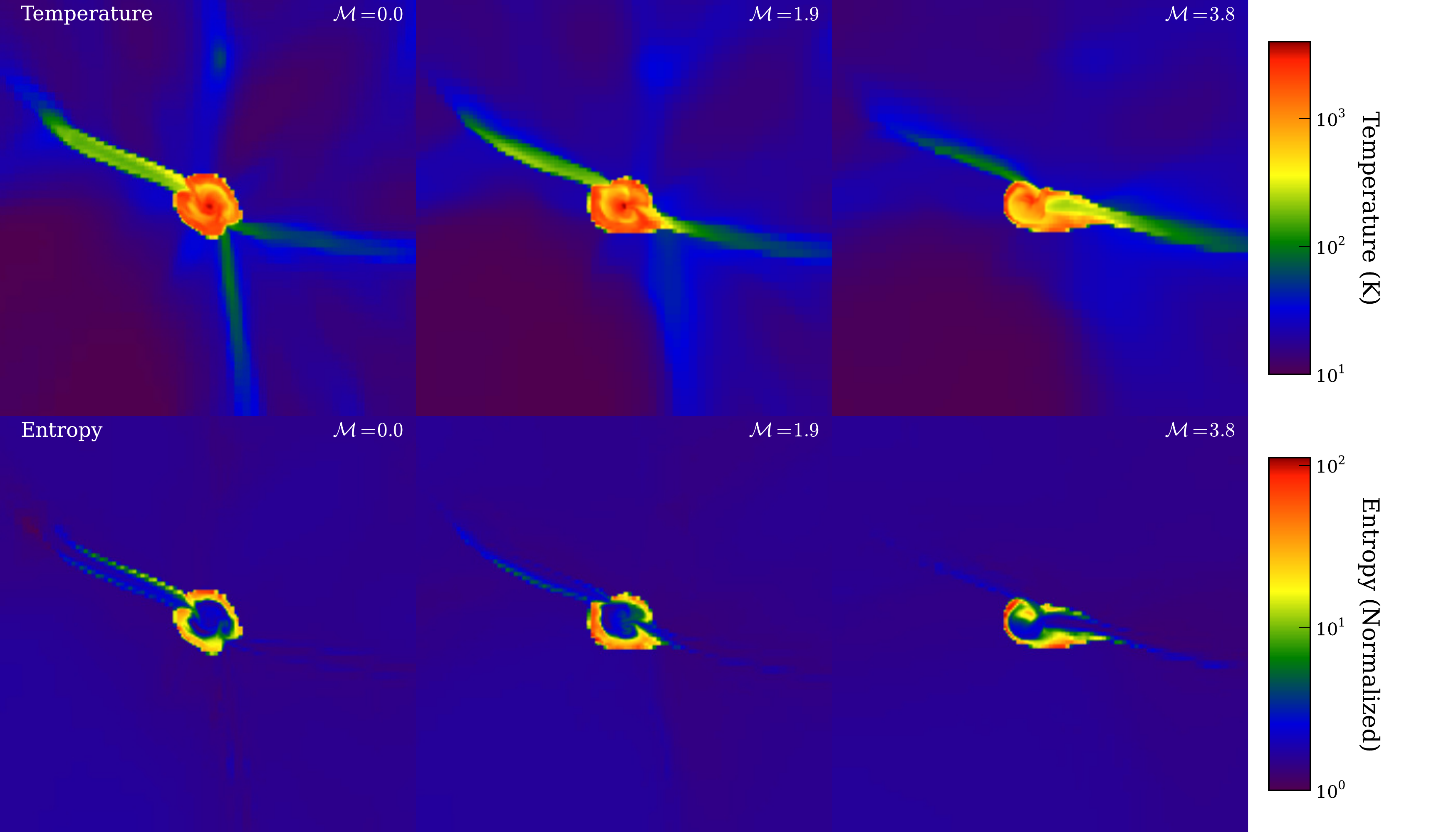}\\
\includegraphics[width=\textwidth]{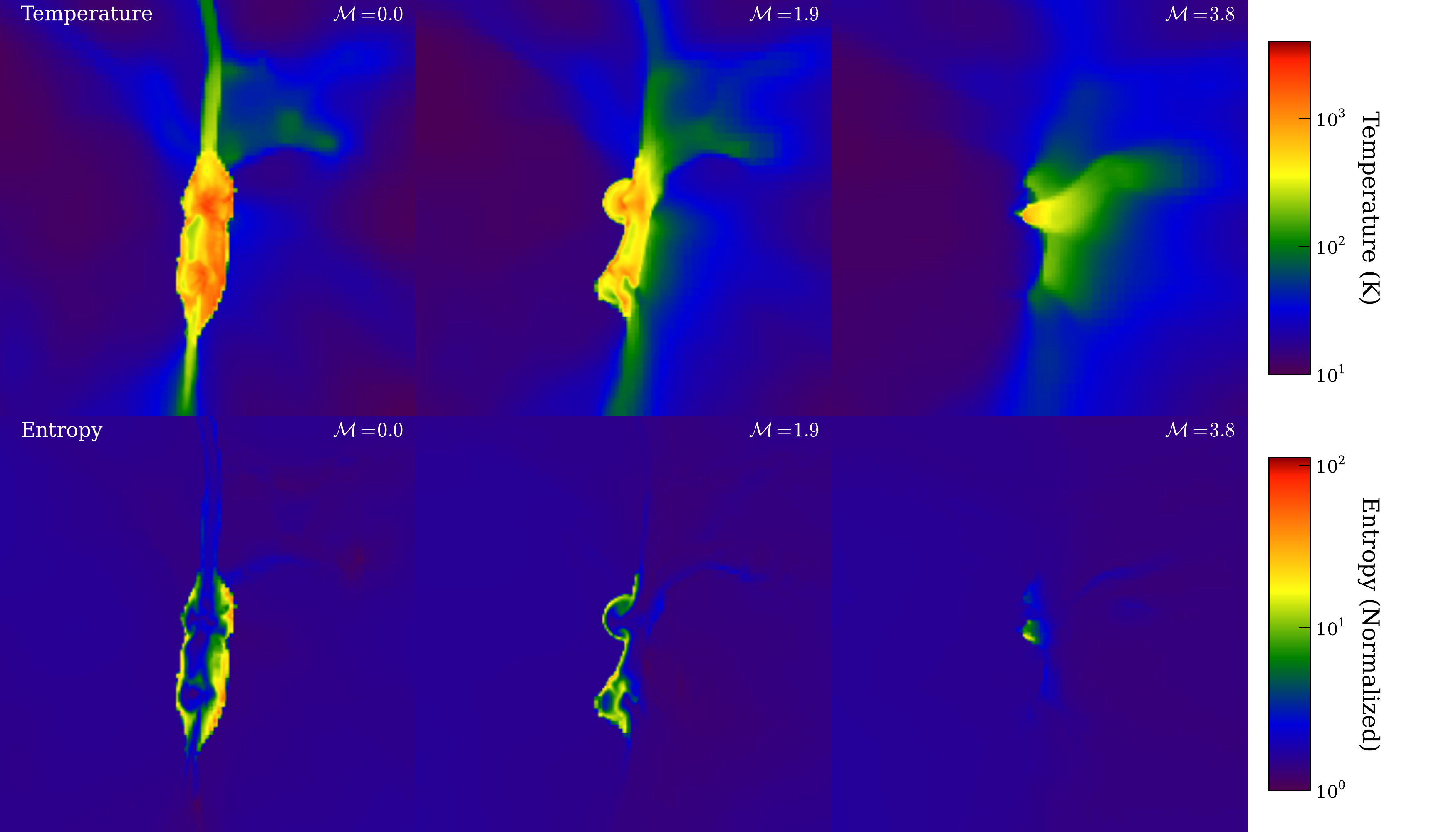}
\end{center}
\caption{Impact of $\vbc$ on shocks in the two most massive halos.  Plotted are  slices of the temperature and the ``entropy,'' which scales as $T/\rho^{2/3}$, of the most massive halo (top) and second most massive halo (bottom) as shown in Figs.~\ref{fig:halo1} and \ref{fig:halo2}, respectively.  The entropy generated from the shocks is confined to the immediate vicinity of the halo.  Note: The frame size shown is a factor of two smaller than in Figs.~\ref{fig:halo1} and \ref{fig:halo2} to show more detail.  \label{fig:entropy}}
\end{figure*}

A supersonic dark matter--baryon velocity difference affects how much gas falls onto each halo as well as the profile of intra-halo gas.  For relatively massive halos ($>10^5~\Msun$), the density of the gas in turn impacts whether they can form enough molecular hydrogen to cool quickly enough and form stars.  However, it is not clear that such massive halos should be significantly impacted by $v_\bc$.  Halos with $>10^5~\Msun$ have circular velocities of $V_{\rm cir}>2.5~$km~s$^{-1}$ at $z=20$, whereas the RMS value for $v_\bc$ is $0.6~$km~s$^{-1}$ at that redshift.  We investigate this question here.  We show that the impact of $v_\bc$ is actually quite significant, even for halos with $V_{\rm cir} = 7~$km~s$^{-1}$ (corresponding to masses of $3\times10^6~\Msun$).  Analytic arguments for why this should be the case are presented in Paper II.

Figures~\ref{fig:halo1}--\ref{fig:halo3} compare density slices of
three individual halos at $z=20$ in simulations with \{$0.5\,\cMpc/h$, $512^3~$initial gaseous resolution elements\}.  The top row of panels use GADGET and the middle Enzo.\footnote{The GADGET slices are generated using the SPH kernel.  The Enzo slices are generated using the yt
  package \citep{yt} and such that the cell thicknesses depends on the level of AMR refinement.}  The bottom row of panels shows the density
  contrast of the baryons to the dark matter or $(1+\delta_b)/(1+\delta_c)$.  Figures~\ref{fig:halo1} and \ref{fig:halo2} show the two largest
halos in these simulations at $z=20$.  Their friends--of--friends group mass in the GADGET simulation with $v_\bc = 0$ are $2\times10^6~\Msun$ and $8\times10^5~\Msun$ (using a linking length of $0.2$).  These masses are slightly smaller in the simulations with nonzero $v_\bc$.  Figure~\ref{fig:halo3} zooms in on a slightly lower mass halo, the $\approx 20^{\rm th}$ most massive halo in the simulation ($m_h \approx
2\times10^5\,\Msun$).  This halo was selected (out of hundreds of other
halos) because its environment
demonstrates a large variety of effects owing to the relative velocity.  

The visual morphology of the virialized region in the most massive halo in the box is not changed significantly between the $\calMbc = 0$ and $\calMbc = 1.9$ cases, but the morphology of this halo for  $\calMbc = 3.8$ is significantly changed (Fig.~\ref{fig:halo1}).  Quantitatively, the central density is decreased by a factor of $2$ between the $\calMbc = 0$ and $\calMbc = 1.9$ cases in Enzo, and by more than a factor of $5$ between the $\calMbc = 0$ and $\calMbc = 3.8$ cases.   In the GADGET simulations, the magnitude of this suppression is similar to that in Enzo, but the central density is always smaller in the GADGET simulations.  
  The differences in central density becomes somewhat larger as the halo mass is lowered (see \S~\ref{ss:sf}).

In addition, the accretion of gas is impacted by $v_{\rm bc}$ for the concentrated filament that forms perpendicular to the bulk flow in Figure~\ref{fig:halo1}.  The orientation with respect to $\bfv_\bc$ of flows onto halos has a notable impact on the amount of gas accreted by the halos in our simulations, and these effects introduce significant stochasticity in the baryonic mass fraction of halos at a fixed dark matter mass. The bottom panel in Figure~\ref{fig:halo1} shows that the dark matter is still present in these filamentary flows, but the baryons are gone, with a slight overdensity downwind.

 The second most massive halo (Fig.~\ref{fig:halo2}) is
 even more altered by the bulk flow of the baryons for
 $\calMbc=1.9$ than the most massive halo.  (Note that there is a neighboring halo that appears in these panels with a somewhat smaller dark matter mass.)  Most striking is the apparent bow shock that develops
 near the virial radius of this halo (and its neighbor) as well as the wisps of downwind gas.
 Remarkably, the halo has little bound gas in the case with $\calMbc = 3.8$.  Again, we see that filamentary structure perpendicular to the baryonic flow is disrupted.  
 
Lastly, Figure~\ref{fig:halo3} shows a halo that is an order of magnitude less massive, with $m_h\approx 10^5~\Msun$.   In this case, Mach cones develop around many of the dark matter overdensities and not just the central halo.  While this halo is just at the mass threshold that is capable of cooling and forming stars \citep{machacek01}, there is no way it can form a star in the $\calMbc = 3.8$ case as the density of the contained gas is only an order of magnitude above the cosmic mean.  

For the case with $\calMbc =0$, the baryons and dark
matter largely trace each other in the simulations in proportion to their cosmological
abundances on scales larger than the Jeans' scale of the gas.
When there is a nonzero velocity difference between the dark matter and baryons, the dark matter and
baryons no longer trace each other even in the most massive halos in our simulations (see the bottom panels in Figs.~\ref{fig:halo1}--\ref{fig:halo3}).   Even the highest density
gas in the most massive halo is offset from the center of the
dark matter halo when $\calMbc = 3.8$.\footnote{We note that because we
  have centered the images on the densest peaks, each frame tends to
  move with the baryon flow as $\calMbc$ increases.}  In general, the gas trails downwind of the dark
matter, especially when the filamentary structure is aligned
perpendicular to the bulk flow of the gas.  Indeed, it is in these
filaments where the gas is often completely swept out of the dark matter well.  The offset of baryonic mass delays accretion as the baryonic wake turns around and falls back onto the halo.  We find that its fallback velocity generally exceeds the typical velocity at which gas is accreted onto a halo in the case $\calMbc =0$, and the fallback of course is lopsided, which further buffets the halo and reshapes the internal gas distribution.  We also found that nonzero $v_{\bc}$ can result in the total angular momentum of the gas changing direction.

\change{
 In Figure~\ref{fig:entropy} investigates how
  $\vbc$ impacts the formation of shocks around the two most massive
  halos in our simulations at $z = 20$. The shocks that heat the gas
  are largely confined to the immediate region around the halos,
  especially when $\vbc > 0$, and only the strongest Mach cones
  generate appreciable entropy.  At $z=25$, when these halos are beginning to virialize, the $\calMbc = 1.8$ halos have bow shocks even before the virial shocks develop in the $\calMbc = 0.0$ halos.

  We have also analyzed the impact $\vbc$ has on the vorticity of halos gas.  Vorticity is defined as $\nabla \times \mathbf{v}_b$, where $\mathbf{v}_b$ is the gas velocity field.  The cosmic initial conditions have zero vorticity, but vorticity is required to seed turbulence and magnetic fields  \cite[see e.g., ][]{1989ApJ...342..650P}, which may alter the properties of the first stars. 
  The primary way to generate cosmological vorticity is via curved
  shocked fronts, and $\vbc$ clearly alters the morphology of shocking
  regions. We found that there was only a modest difference in the
  magnitude of voriticity with $\vbc>0$ in the three most massive
  halos at $z=20$, when most of the gas had already decelerated.
  However, at $z=25$ the most massive halos in simulations with $\calMbc =
  1.8$ had an order of magnitude more vorticity compared to those in that with $\calMbc =
  0.0$.  At $z=20$, the
 $\sim 10^5~\Msun$ halos in the simulations had higher vorticities than in the case with $\vbc
  >0$.  (Note that while cosmic vorticity should be conserved, it is almost certainly damped out with time by the high artificial/numeric viscosity in our simulations.) These trends suggest that for $\vbc > 0$, halos develop shocks
  earlier and have greater vorticity (and, thus, are more turbulent) than halos in previous simulations of
  the first halos and stars with $\vbc = 0$.}

Overall, the GADGET and Enzo simulations show a remarkable level of
agreement on the structure and morphology of the three halos in
Figures~\ref{fig:halo1}--\ref{fig:halo3}.  The location and shape of
the gas and dark matter structures are similar in both codes'
simulations on scales as large as the box size ($\sim1 ~\cMpc$;
Fig.~\ref{fig:images}) to as small as a tenth of the virial radius of
$10^5~\Msun$ halos ($\sim 0.1~$kpc; Fig~\ref{fig:halo2}).  The halos
in GADGET tend to be less dense than the halos in Enzo. \change{One
  possible cause of this may be the different effective resolutions
  between two codes:} The \change{minimum} softening length in GADGET
at $z=20$ is $0.04~$kpc (approximately two decades smaller than the
halo diameter), and the AMR refinement criteria that was used for Enzo
is super-Lagrangian.  \change{However, in testing Enzo simulations
  with smaller boxes and more aggressive refinement, we found that the
  earlier the refinement was initialized, the more concentrated the
  halos appeared at later times.}  Note that none of the simulations
includes molecular hydrogen cooling and so this comparison is not
sensitive to cooling rates. \change{ The simulations also have similar
  thermal properties throughout the volume, again with the exception
  of the peak densities found in the halos.  We refer the reader to
  Appendix~\ref{ss:temp} to directly compare the thermal propertiers
  of the gas in the various simulations, and the impact particle
  coupling has on the GADGET simulations. }

\subsection{The First Structures}
\label{ss:firststructures}

\begin{figure}
\begin{center}
\includegraphics[width=\columnwidth]{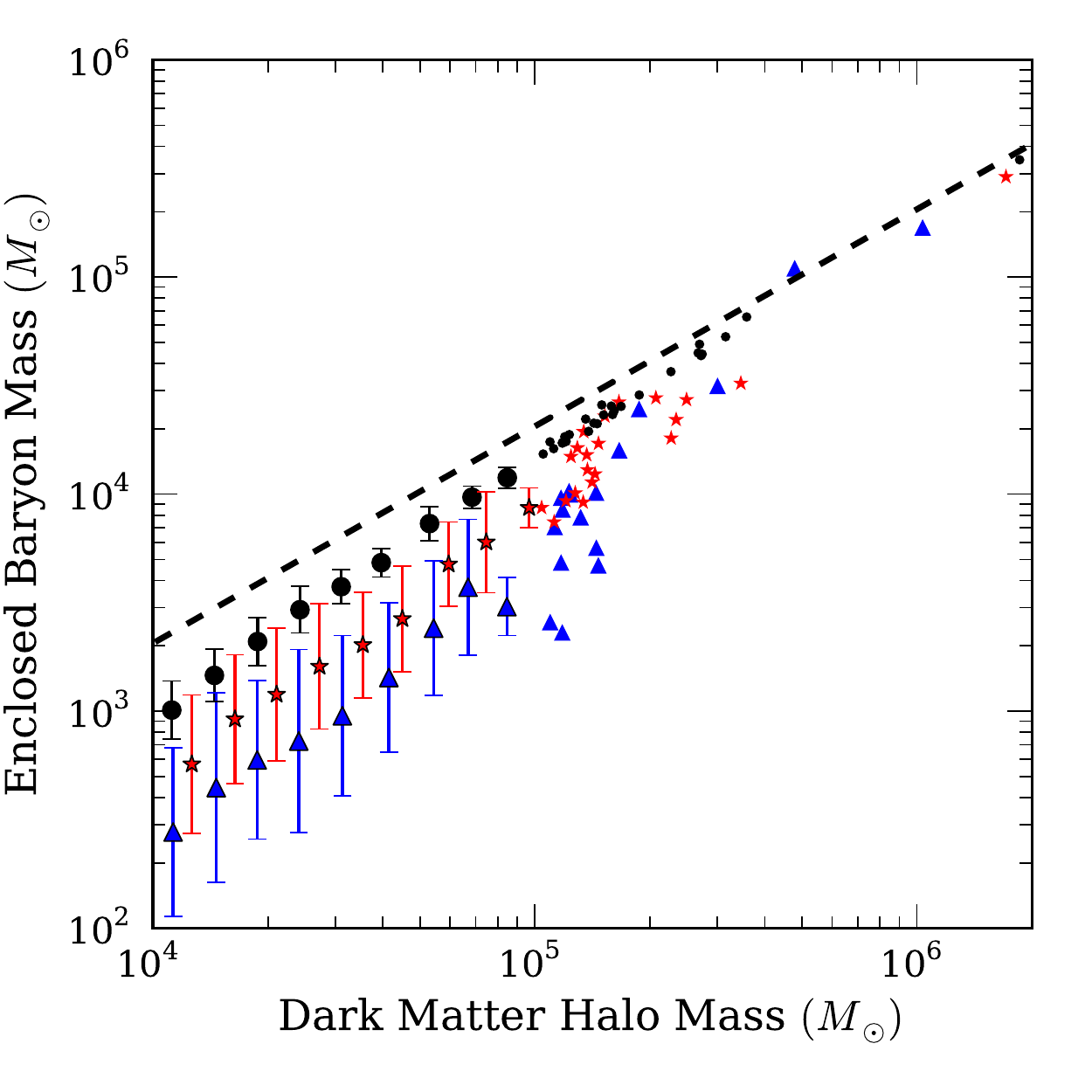}
\end{center}
\caption{Mass of baryons in halos as a function of halo mass, $m_h$.
 For each halo, we find the peak
  gas density and, then, determine the mass of baryons enclosed within
  $r_h$, the radius that encloses $m_h$ of dark matter.  We describe the reasons behind this nontraditional setup in the text.  The black
  circles, red stars, and blue triangles represent halos from Enzo simulations with $0.5~\cMpc/h$, $512^3~$initial gaseous resolution elements and for the cases
  $\calMbc = 0.0$, $1.9$, and $3.8$, respectively.  For $m_h > 10^5\,\Msun$, we show each individual halo.  The remaining data at  $m_h < 10^5\,\Msun$ show the mean and standard deviation of the \emph{logarithm} of the values for individual halos.  The dashed line equals $\Omega_b/\Omega_c \,m_h$.
  \label{fig:gasmassfraction}}

\end{figure}

As is evident from Figures \ref{fig:halo1}, \ref{fig:halo2}, and \ref{fig:halo3}, the infall of gas
onto the first halos is suppressed in the $\calMbc >0$ simulations relative to the $\calMbc =0$ runs for typical streaming velocities \citep{stacy11,greif11,fialkov11}.  In Figure~\ref{fig:gasmassfraction}, we investigate the amount of mass in baryons that accretes onto
each halo as a function of the dark matter halo mass for Enzo simulations with $0.5~\cMpc/h$, $512^3~$initial gaseous resolution elements.\footnote{The dark matter mass can be decreased
  by as much as a factor of two for the case $\calMbc=3.8$, even in the largest halos. In Paper II, we discuss in detail the suppression of the dark matter
  mass function for the largest GADGET simulations when $\calMbc = 0$
  and $1.9$.}  Below $10^5~\Msun$ in this plot, we instead show the logarithmic mean of the baryonic mass and its variance for the halos within a mass bin.  (The physical mean of these curves is similar to the logarithmic mean.)  There is no well defined way to measure the baryonic mass contained within the
first structures, especially when $\calMbc > 0$ so that the dark
matter and baryons often do not trace one another.  In many instances (even for the most massive halos in our simulations), the gas and dark matter density
peaks are offset by as much as $r_{200}$ -- the distance from the halo density peak at which the the dark matter overdensity falls below $200$ -- so that it is unclear what is the best method to center when reporting spherically
averaged quantities.  We choose to center on the gas density peaks in this study.  To find these peaks, we first search for the densest
peak within two viral radii of the dark matter center of mass, and
then we select a sphere with radius $r_h$, such that the total
enclosed dark matter mass is $m_h$, as found by the HOP halo finder
\citep{HOPfinder}.  While this algorithm is non-traditional, it yields a conservative estimate for the amount of mass that is contained.  For example, we find that if we use a criterion more similar to that used in other analyses such as \citet{naoz11} (in which we start with the dark matter density peak or dark matter center of mass, or especially if we calculate the gas within
$r_{200}$ rather than $r_h$), the resulting baryonic masses are significantly lower for halos in simulations with $\calMbc > 0$.  In addition, the halo-to-halo scatter also increases.  For example, if we center on the dark matter center of mass, as opposed to gas density peak, and calculate the gas within a distance $r_h$, this decreases the baryon mass by
$0.94\pm 0.16, 0.73\pm 0.38,$ and $0.50\pm 0.36$ on average ($\pm$ the standard deviation of the scatter) for
$\calMbc = 0.0, 1.9,$ and $3.8$, respectively.  Finally, we find consistent results
between the Enzo and GADGET simulations using a similar gas-finding algorithm.

Figure~\ref{fig:gasmassfraction} illustrates that increasing the value of $\calM_\bc$ in the simulations significantly reduces the mass of baryons that have
accumulated into dark matter halos.  To a lesser extent, it also lowers the mass of dark matter that has
collapsed into the most massive halos.  In addition, the scatter in the
baryonic mass at fixed halo mass increases drastically with $\calM_\bc$.  For a $10^5\,\Msun$ halo and
$\calMbc = 3.8$, the baryon fraction is
approximately half that found when there is no relatively velocity.
The halo-to-halo scatter in the baryonic fraction, on the other hand, increases to $100\,\%$ from only
$10\,\%$  between $\calM_\bc = 0$ and $\calM_\bc = 3.8$.  For cases in which there is a bulk
velocity between the dark matter and baryons, the enclosed baryon mass
in individual halos becomes dependent on the environment of
the halo.  From our earlier investigations of individual halos, it is clear that halos
found in filaments perpendicular to the bulk flow appear more
affected, as do halos without any apparent nearby structure.  Such environmental dependences are likely sourcing the added stochasticity as $v_\bc$ is increased.

\subsection{Star Formation}
\label{ss:sf}

As we described in \S~\ref{ss:firststructures}, the baryon-dark matter
velocity difference suppresses the accretion of gas into the first
halos, thereby reducing the density and temperature of these
structures. This suppressed accretion also strongly damps the rate of star formation in
the first halos.  Our simulations do not follow the gas to densities where it becomes a star since we do not have molecular hydrogen cooling.  Thus, we use the mass of gas that
can cool in a Hubble time from molecular hydrogen cooling as a proxy
for the star formation rate in our simulations.   However, we find that our relative comparisons among the simulations do not change if we instead take the gas mass that can form within $0.1$ Hubble times.    In our opinion, it is not necessarily a disadvantage to use our cooling criteria as a proxy for star formation rather than following the cooling and condensing gas to much higher densities:  Firstly, simulations that follow gas to much higher densities do not all agree on the character of star formation in the first halos (e.g., \citealt{greif11b}).  In addition, feedback processes either from stellar HII regions within the halo \citep{alvarez06, yoshida07} or the cosmological Lyman-Werner background \citep{haiman00, machacek01} drastically increase the complexity of modeling such star formation beyond when the first star has formed in the halo or when the cosmological star formation rate density has reached a value of $\sim 10^{-6}~\Msun~{\rm yr}~\cMpc^{-3}$.

\begin{figure}
\begin{center}
\includegraphics[width=\columnwidth]{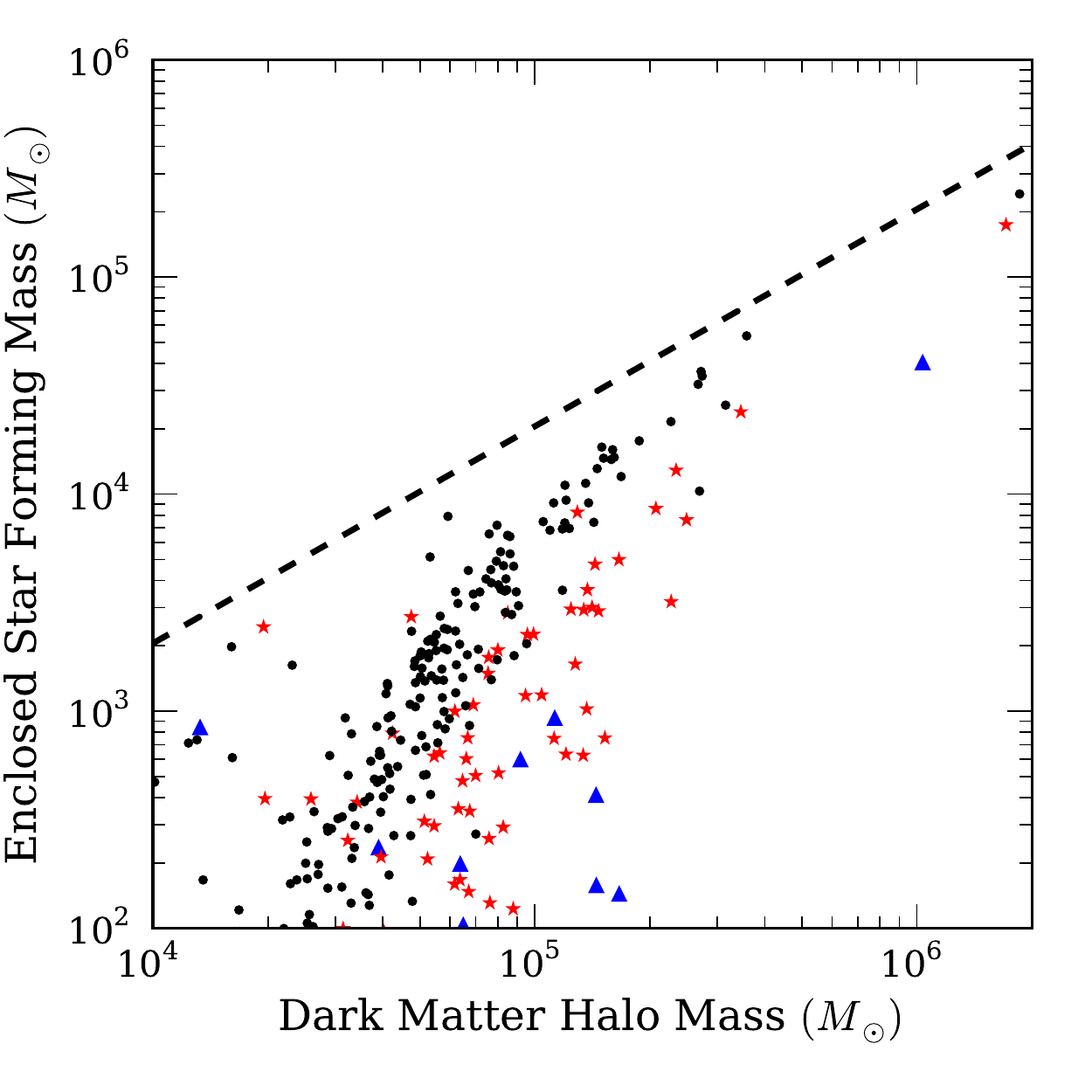}
\end{center}
\caption{ Mass of baryons in halos that can cool in a Hubble time as a function of halo mass in the $0.5~\cMpc/$ Enzo AMR simulations.
  For each halo, we find the mass of
  baryons enclosed in $r_h$ that can cool in a Hubble time, where $r_h$ is the radius that encloses $m_h$ of dark
  matter.  The black circles, red stars, and blue triangles represent data from the
  $\calMbc = 0.0$, $1.9$, and $3.8$ respectively.  The dashed line equals $\Omega_b/\Omega_c m_h$. \label{fig:sfmassfunction}}
\end{figure}

\begin{figure}
\begin{center}
\includegraphics[width=\columnwidth]{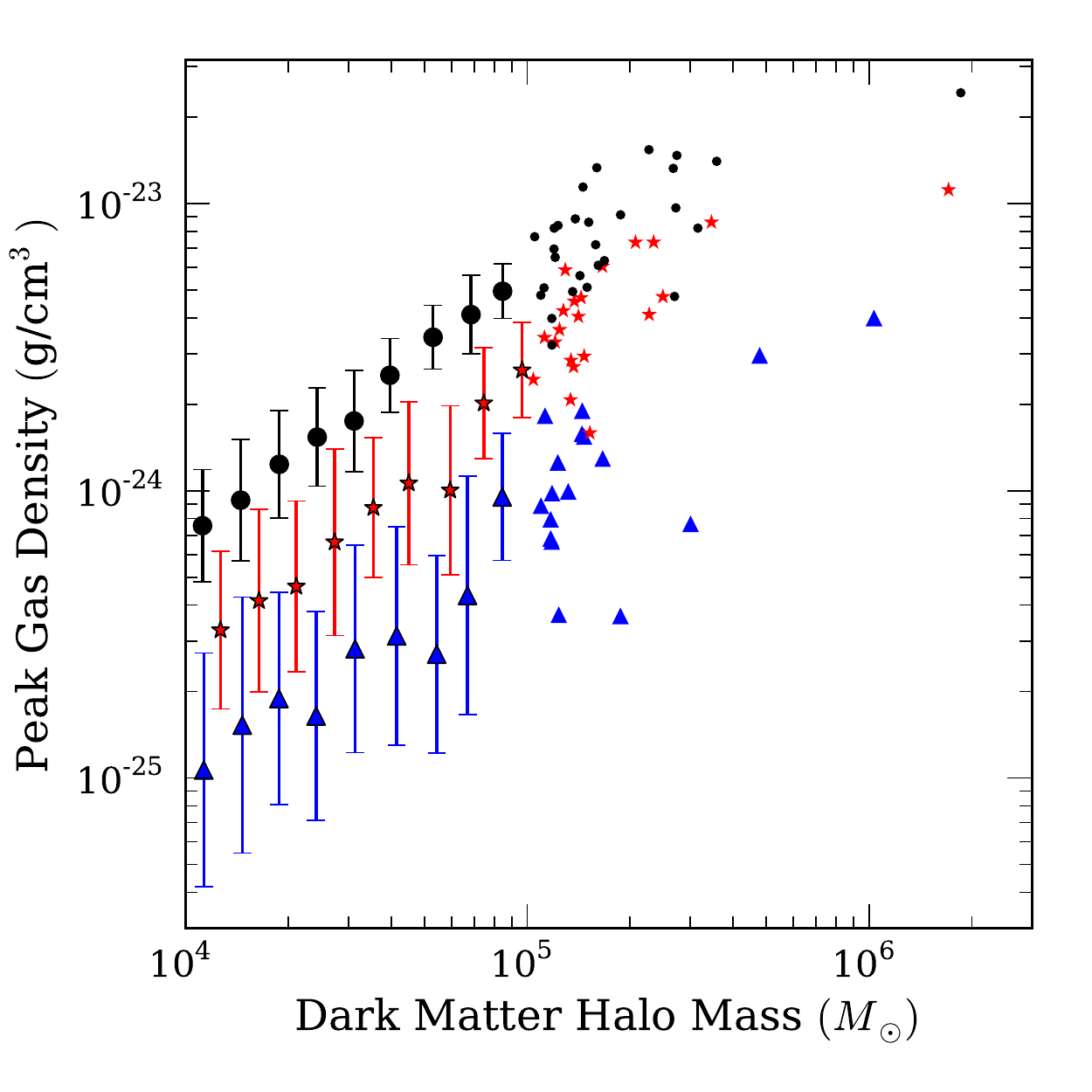}
\end{center}
\caption{Peak gas density in halos as a function of mass in the $0.5~\cMpc/h$, $512^3~$base grid Enzo AMR simulations.
  For each halo, we find the peak
  gas density within two viral radii of the halo center of mass.  The black
  circles, red stars, and blue triangles represent data from the
  $\calMbc = 0.0$, $1.9$, and $3.8$ simulations, respectively.  For $m_h > 10^5\,\Msun$, we show the peak density for each individual halo.  The remaining points with error bars shows the mean and standard deviation of the {\em logarithm} of the values.  For a given dark matter halo mass, the peak density is reduced by approximately half ($\approx 1/8$) when $\calMbc = 1.9$ (3.8) as compared to $\calMbc = 0.0$. At $z=20$, the mean gas density is $4\times10^{-27}\,$g/cm$^{-3}$. \label{fig:plotmax}}
\end{figure}

Figure~\ref{fig:sfmassfunction} shows the mass in baryons that
have cooling times less than a Hubble time as a function of the halo
mass for our Enzo simulations with boxsize $0.5\,\cMpc/h$ and $512^3$ initial gaseous resolution elements.  To calculate the cooling time, we use the formula in \citet{tegmark97} under the crude approximation that $\log(1+ N_{\rm rec}) =1$ to calculate the amount of molecular hydrogen, where $N_{\rm rec}$ is the number of recombinations.  This equality approximately holds for gas at the virial density of halos at $z=20$, and it allows us to calculate the cooling rate from a single time slice rather than following the density evolution of a fluid element.  To tabulate the amount of cooling gas, we search within a sphere of $r_h$ from the halo center of mass.  However, we find that our qualitative results are insensitive to this choice and searching within $r_{200}$ around the gas density peak yields similar numbers.

 We find that the dark matter mass threshold required to have high enough temperatures
and densities to cool increases with $\calMbc$ in agreement with the picture in previous studies \citep{greif11, stacy11, fialkov11}.  Surprisingly, we find
that even some of the most massive halos in our boxes (up to $10^7~\Msun$ in the $1~$Mpc boxes reported in Paper II) have less
cooling gas for the case with $\calM_\bc =1.9$ than with $\calM_\bc=0$, in agreement with \citet{naoz11}.  The amount by which the star formation rate is decreased has significant variance at a single mass.  On a
halo-by-halo basis $\{98\%$, ~$57\%$,~ $11\%\}$ of halos with dark
matter halo masses $m_h>4\times 10^4\,\Msun$, have more than
$100\,\Msun$ of gas that will cool in a Hubble time for $\calMbc =
\{0.0,~1.9,~3.8\}$.  Even for $m_h>1\times 10^5\,\Msun$, $\approx 2/3$ of halos in the $\calMbc =3.8$ simulation have less than $10^2~\Msun$ in gas that can cool.    We just note here that Figure~\ref{fig:sfmassfunction}
uses our Enzo simulations, with the HOP halo finder to determine dark matter
halo sizes, which we found to be reduced compared to the dark matter
halos in GADGET.  

The
impact $\calMbc$ has on the peak gas density in halos is related to the impact $\calMbc$ has on the amount of gas that can cool.  In
Figure~\ref{fig:plotmax}, we show the peak gas density in each halo  in the $0.5~\cMpc/h$, $512^3~$base grid Enzo AMR simulations as
a function of the halo dark matter halo mass. 
  We find that the peak central density decreases by
nearly an order of magnitude, on average, when $\calMbc = 3.8$.  The
dispersion of peak densities also increases significantly, similar to
what we found in the baryon masses of the halos.

In Paper II, we exclusively use the GADGET simulation on a larger box
to look at how the total star formation rate is modulated by
$\calMbc$.  In the smaller box sizes reported here, we find that Enzo,
on average, has higher gas densities and temperatures in the halos
(see our discussion in \SS~\ref{ss:temp} and
Fig.~\ref{fig:phase}). This results in approximately four times as
much gas that can cool to form stars.  In addition, the number of
halos that form stars is approximately twice as large as found in
GADGET.  However, the fraction that star formation
is reduced by $\calMbc$ is the same in both
codes.  %

As a final aside, given the large offset between the baryons and dark
matter in our simulations, it seems improbable that the first stars
ever form at the dark matter density peak in the halo.  Thus, we find
it unlikely that stars supported by dark matter annihilation
\citep{2008PhRvL.100e1101S} ever existed, even with favorable dark
matter annihilation cross sections.

\section{conclusions}

This study investigated how gas was captured into the first structures during the cosmic Dark Ages.  This capture is relevant to the formation of the first stars and the high-redshift 21cm signal.  In the concordance $\Lambda$CDM cosmological model with nearly scale-invariant primordial potential perturbations, the initial conditions that set the distribution and velocities of matter are well constrained by observations of the CMB and of large-scale structure.  This study aimed to follow the evolution of these (initially small) matter fluctuations into their nonlinear state at lower redshifts.  However, there are challenges with achieving this objective.  A simulation must have the resolution to capture scales below the Jeans'
scale of the gas as well as be large enough to capture a representative sample of the Universe.  We discussed the specifications necessary to meet these competing objectives.  

We ran a large number of simulations of the early Universe with both the GADGET and Enzo cosmology codes.
Our simulations are the first to initialize the gas and dark matter
self-consistently from linear theory on scales where gas pressure is
important and at redshifts where the baryons and dark matter do not
trace each other.  There is no physical or significant computational reason to not
initialize cosmological gas+dark matter simulations with anything but the
full linear solutions, but this appears to not have been done previously.  In
contrast to previous numerical studies, we initialized most of our
simulations in a manner that included (1) the impact of pressure on
the growth and rate of growth of modes, (2) temperature fluctuations,
(3) the correct baryonic and dark matter velocities, and (4) transfer
functions that account for the dark matter-baryon velocity
differential (under the approximation that Compton drag from the CMB was zero at $z<1000$).  We showed that simply boosting $v_\bc$ at
the initial redshift of the simulation, as done in prior studies, misses much
of the impact of $v_\bc$ on the linear growth of density fluctuations.

Because our simulations were run with two very different codes (the
tree--particle mesh plus SPH code GADGET and the nested particle mesh plus AMR code Enzo), the comparison of these codes' simulations provided a
relatively controlled test of the codes themselves.  Despite their very different algorithms, the
$\sim 10^6~\Msun$ halos that formed in GADGET and Enzo (with AMR) simulations have
similar appearances and their thermal properties agree remarkably
well.  While the densities in these halos were lower in GADGET, we found no evidence that GADGET either significantly under-
or over-estimates the entropy produced in weak shocks from structure
formation despite its SPH hydro-solver.  The largest difference
between the simulations with the two codes was from spurious gas--dark matter particle
coupling in the GADGET simulations, \change{which was most pronounced} when $v_\bc = 0$. 
%
 We found that this coupling is slightly
smaller when we use the common prescription of staggering gas and dark
matter particles at half an interparticle spacing (but with the same
glass file) as opposed to using two separate glass files for the gas and dark matter.  This coupling is likely to have impacted all previous cosmological SPH simulations. \change{However, by using adaptive gravitational softening for the baryons in the simulations, we were able to eliminate the bulk of this particle coupling.  We recommend that all cosmological SPH simulations use adaptive gravitational smoothing lengths, as the effects of particle coupling evade standard convergence tests.} \\

Our primary focus was to model the impact of $v_\bc$ during the cosmic Dark Ages on the formation of the first nonlinear structures, especially the first stars.  To do so, we used a suite
of $\sim 20$ GADGET and Enzo simulations to explore the formation of the first
structures in the Universe as well as the thermal evolution of the gas.  We found
that:
\begin{enumerate}
\item The halo gas in these simulations is often found
significantly downwind and with lower densities in the simulations with $\calM_\bc > 0$.  The lower densities delay the formation of the first stars.  This delay is consistent with what was found in previous studies \citep{stacy11, greif11}.  These effects impact the formation of the first stars and could imprint fluctuations in high-redshift 21cm backgrounds (as discussed in Paper II).  The gas that accumulates in the halos is off-center from the dark matter halo, with
the density peaks in the gas and dark matter offset by as much as
$r_{200}$ for the case $\calM_\bc = 2$ (which is near the cosmological average of $\calM_\bc$).  Furthermore, we find that the maximum gas density can be suppressed on average by roughly an order of magnitude between the cases $\calM_\bc = 0$ and $\calM_\bc = 4$ for halos with masses of $10^4~\Msun$ and even up to $10^6~\Msun$.
\item Dynamical friction induced by the gas flowing by the
dark matter halos induces visible Mach cones, and the tug from the mass in these supersonic wakes acts to erase the velocity difference between the dark matter and baryons.  We showed that the dynamical friction timescale for $\gtrsim 10^4~\Msun$ halos is less than a Hubble time for typical $\calM_\bc$.  Most of the overdense gas in our simulation has decelerated into the dark matter frame via this process by $z =20$.  These downwind wakes eventually crash back onto dark matter halos, with larger infall velocities than most of the accreted gas.  However, we found that these shocks did not significantly heat the IGM. 
\item The halo environment plays a significant role in how much gas makes it into a halo in cases with typical $\calM_\bc$.   For example, we find that dark matter filaments perpendicular to the gas flow are often
devoid of gas.  Whether a halo has gas or not depends on the orientation of intersecting filaments.  As a result, the baryonic mass fraction of halos (as well as the fraction of mass that can cool and form stars) shows significant stochasticity at fixed halo mass in the simulations with $\calM_\bc > 0$.  
\end{enumerate}


  We would especially like to thank Dusan Keres and Mike Kuhlen for their help with GADGET and Enzo.  We thank Gianni Bernardi, Lincoln Greenhill, and Martin White for useful discussions, and Rennan Barkana and Smadar Naoz for helpful comments on the manuscript.  Computations described in this work were performed using the GADGET3 and Enzo codes, and with the yt analysis software \citep{yt}.  GADGET3 was developed by Volker Springel, and Enzo was developed by the Laboratory for Computational Astrophysics at the University of California in San Diego (\url{http://lca.ucsd.edu}).  RO and MM are supported by the National Aeronautics and Space
Administration through Einstein Postdoctoral Fellowship Award Number
 PF0-110078 (RO) and PF9-00065 (MM) issued by the Chandra X-ray Observatory Center, which is
operated by the Smithsonian Astrophysical Observatory for and on
behalf of the National Aeronautics Space Administration under contract
NAS8-03060. This research was supported in part by the National Science Foundation through TeraGrid resources provided by the San Diego Supercomputing Center (SDSC) \citep{teragrid} and through award number AST/1106059.

\appendix

\bibliographystyle{apj}
\bibliography{dynamical_heating}

\begin{appendix}

\section{A. linear theory}
\label{ap:linth}
The system of equations for the linear evolution of the matter in the presence of an initially uniform velocity difference between the dark matter and baryons is \citep{tseliakhovich10}
\begin{eqnarray}
\dot{\delta}_c &=&  - \theta_c \label{eqn:lintheory1},\\
\dot{\theta}_c & = & - 3 H^2/2 \, (\Omega_c \delta_c + \Omega_b \delta_b) - 2H \theta_c,\\
\dot{\delta}_b &=&  - i \,a^{-1} \, \bfv_{\bc} \cdot \bfk \, \delta_b - \theta_b, \\
\dot{\theta}_b & = & - i \, a^{-1} \, \bfv_{\bc} \cdot \bfk \, \theta_c  - 3 H^2/2 \, (\Omega_c \delta_c + \Omega_b \delta_b) - 2H \theta_b + c_s^2 \, k^2 a^{-2} \delta_b,
\label{eqn:lintheory4}
\end{eqnarray}
where $\theta_X$ is related to the velocity field via $\bfv_X = -i a k^{-2} \bfk  \, \theta_X$ (since $v_X$ has zero vorticity until the onset of non-linearity for inflationary initial conditions), and $c$ denotes cold dark matter and $b$ baryonic material.  These equations use that the homogeneous component of $v_\bc$ redshifts away as $v_\bc \propto (1+z)$.
Equations (\ref{eqn:lintheory1} - \ref{eqn:lintheory4}) use a similar notation to that in \citealt{tseliakhovich10} except that they are in the frame of the dark matter (which results in less oscillatory solutions in imaginary space once the baryons begin to fall into the dark matter potential wells).  An additional equation for the temperature is required to solve for $c_s$.  We use the temperature equation in \citet{naoz07}, which includes adiabatic processes and Compton heating off of the CMB.

\section{B. Particle coupling}
\label{sec:partcoup}
\change{During the initial comparison of the GADGET and Enzo
  simulations, we found a significant excess in power in the gas
  density power spectrum on small scales in our GADGET simulations relative to Enzo (and to linear theory), especially in the GADGET simulations with  $\vbc=0$ (see Fig.~\ref{fig:ltcomp}).  This excess power remained with the same magnitude
  and slope even when the particle resolution was increased.  Appendix~\ref{ss:coup} describes how this excess owes to particle
  coupling, and Appendix~\ref{ss:temp} shows how
  the coupling alters the distribution of temperatures and densities in our GADGET simulations.  For the simulations discussed in the body of this paper, we used a variable gravitational softening length (with this set equal to the softening length), which successfully removes at least the bulk of this coupling.  Variable softening is not typical in such simulations, but should be done in all further cosmological SPH simulations.  It simply involves turning on the compile-time flags ADAPTIVE\_GRAVSOFT\_FORGAS  (and also ADAPTIVE\_GRAVSOFT\_FORGAS\_HSML in GADGET3).}

\subsection{B.1 Origin of Coupling}
\label{ss:coup}
Here we attempt to understand how particle coupling should scale with redshift, box size, and particle number.  This boils down to comparing the escape velocity of particles from each other's potential well to their relative velocity.  The escape velocity at a proper distance $r$ from a dark matter particle of mass $M_{p}$ is
\begin{equation}
v_{\rm esc, p}  \equiv \left(\frac{2 \,G \, M_p}{r}\right)^{1/2} =   0.63 \, \left(\frac{M_p}{5 \, \Msun} \right)^{1/3} \left(\frac{z}{200} \right)^{1/2} \left( \frac{f_{\rm ms}}{0.04}\right)^{-1/2}~~~{\rm km~s^{-1}},
\end{equation}
where $f_{\rm ms}$ is $r$ in units of mean interparticle spacings and note that $5~\Msun$ is $M_p$ in the $0.2~\cMpc/h$, $2\times 512^3$ particle GADGET simulations.  The gravitational force increases down to the softening length, which we have taken to be $f_{\rm ms}=0.04$ in our GADGET simulations. (The rule of thumb for cosmological simulations is $0.03 -- 0.04$.)

Compare $v_{\rm esc, p}$ to the relative velocity of particles.
The Hubble flow dominates over peculiar velocities in determining the relative velocity of neighboring particles.  The Hubble velocity of particles during matter domination is
\begin{equation}
v_H = H \, r  =  0.011 \, \left(\frac{M_p}{5 \, \Msun} \right)^{1/3} \left( \frac{z}{200} \right)^{1/2} \left( \frac{f_{\rm ms}}{0.04} \right)~~~{\rm km~s^{-1}}.
\end{equation}
Both $v_{\rm esc, p}$ and $v_H$ have the same scaling with redshift and $M_p$.  Thus, box size and particle number do not change the typical amount of coupling between a particle and its neighbors, explaining the trends seen in our numerical tests.  Also, $v_{\rm esc, p}$ and $v_H$ are only comparable at approximately one interparticle separation.  At smaller separations, $v_{\rm esc, p}$ is larger, meaning that particles are likely to become trapped in the potential well of other particles. 

However, the velocity of the baryons relative to the dark matter naturally \change{alleviates {\em some}} of this coupling as $v_\bc \sim 5 \times (z/200)~{\rm km~s^{-1}}$, so that the energy of gas particles is such that they can travel in and out of the potential wells of the dark matter particles. \change{However even in the absence of dynamical friction, the different scaling with $z$ between the relative velocity and the escape velocity ensures that even in the case with $v_\bc > 0$, some particle coupling is unavoidable with a fixed smoothing length.  
}

\subsection{B.2  Impact on Temperatures and Densities in the Simulations}
\label{ss:temp}
\begin{figure*}
\begin{center}
\includegraphics[width=\textwidth]{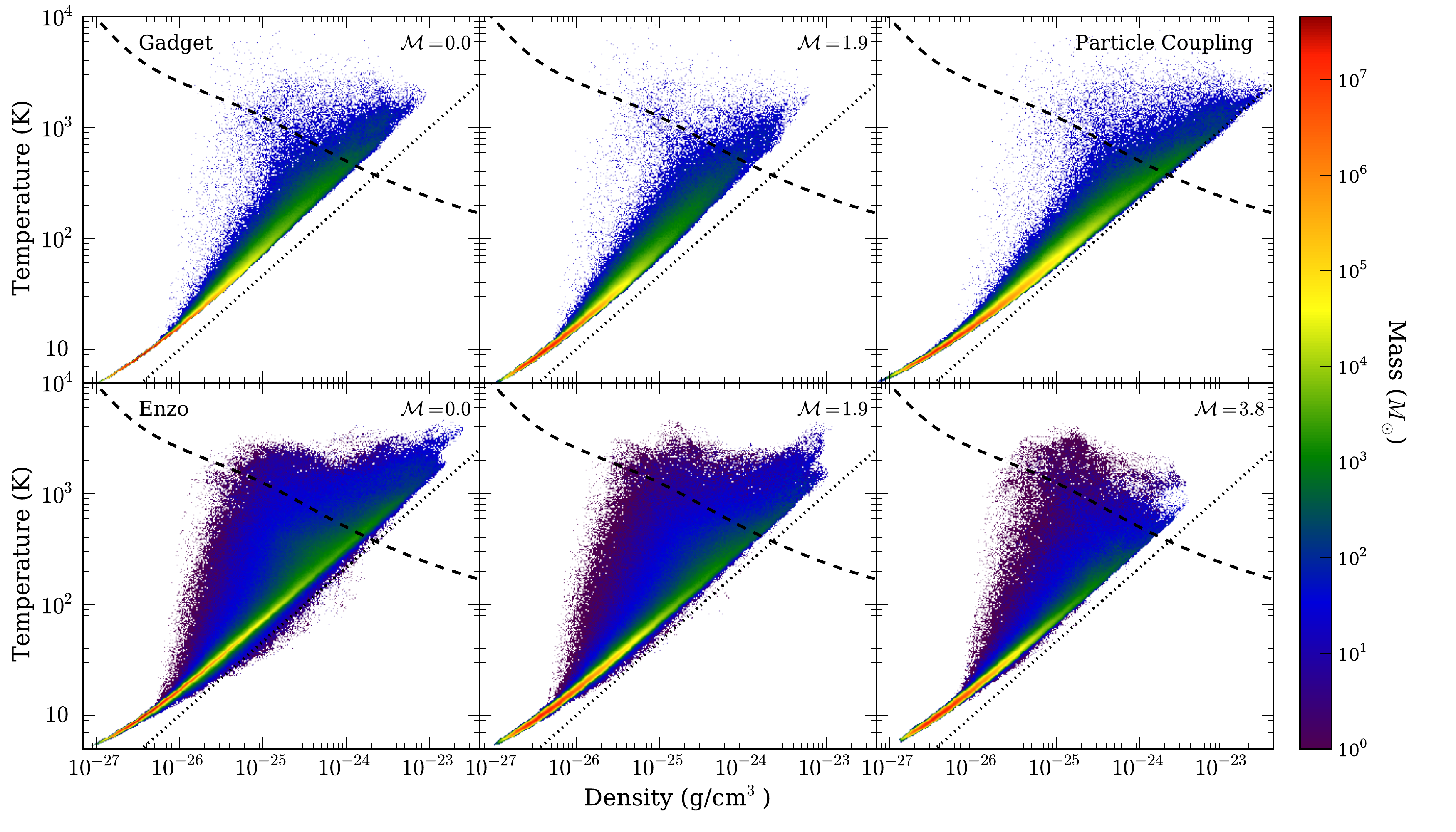}
\end{center}
\caption{Temperature--density phase diagram of the $z=20$ gas in the $0.5~\cMpc/h$, $2\times512^3$~base-grid resolution element simulations.  The top
  (bottom) panel shows a two-dimensional histograph of gas mass as
  a function of temperature and density for the GADGET (Enzo) simulations with $\calMbc
  = 0.0, 1.9,$ and $3.8$ (from left to right, respectively). \change{The top right panel should be compared to the top left panel. It shows the effect of particle coupling on the phase diagram from the GADGET simulation that used a fixed softening length with $\calMbc = 0.0$. }  For reference, the region
  above the dashed lines can cool in a Hubble time owing to molecular
  hydrogen (see the text for details), and the dotted lines show a single arbitrarily normalized adiabat (temperature $\propto \rho^{2/3}$).   Each point is color coded to represent the mass that falls in one of $400\times400$
  bins in log space.  GADGET contains none of the lowest mass, purple points because 
  it has a gas particle mass of $16.75\,\Msun$. \label{fig:phase}}
\end{figure*}

\begin{figure*}
\begin{center}
\includegraphics[width=8cm]{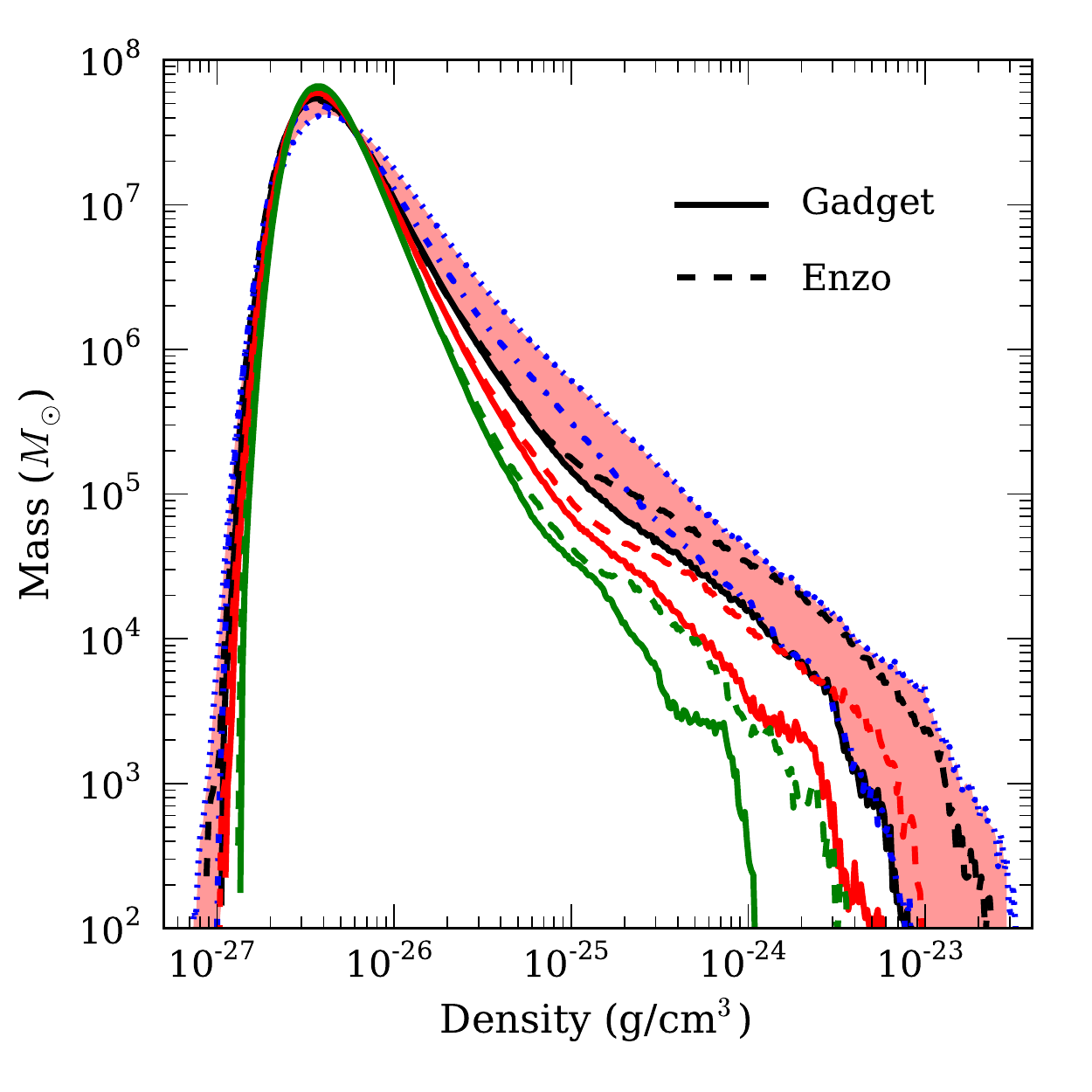}
\includegraphics[width=8cm]{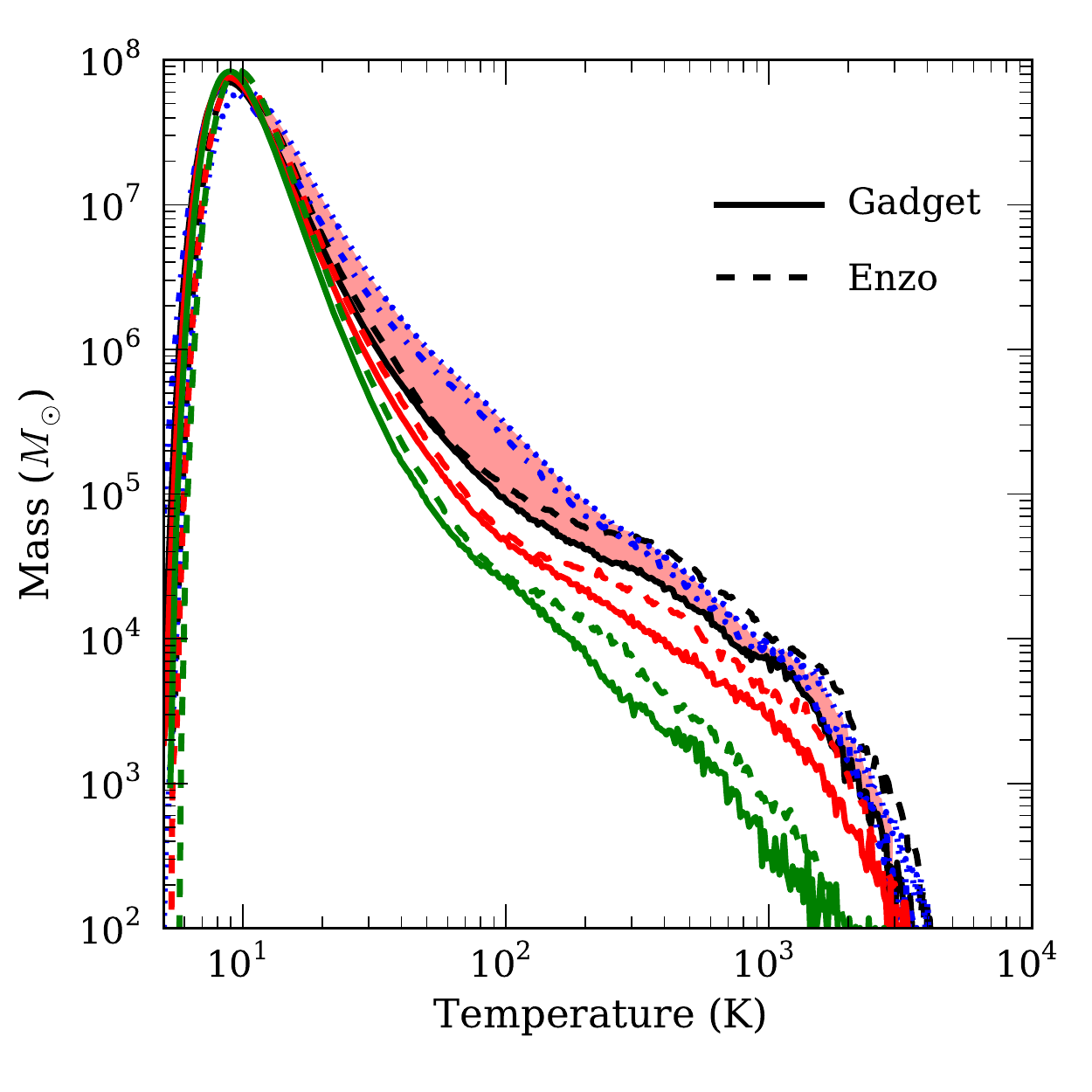}
\end{center}
\caption{Histograms of the density (left panel) and temperature (right
  panel). In both panels, we plot mass--weighted PDFs with solid (dashed) curves from our $0.5~\cMpc/h$ GADGET
  (Enzo) simulations.  The \{black, red, green\} curves correspond to $\calMbc
  = \{0.0,1.9,3.8\}$ (roughly from top to bottom).  The density
  histograms for Enzo and GADGET agree rather well \change{at
  $\rho_b \lesssim 10^{-25}$g/cm$^3$.    The blue dotted curves show the results of our simulations with $\calMbc = 0.0$ and a fixed gravitational softening length.  We attribute the discrepancy between these curves and the other GADGET curves (solid black) to gas-dark matter particle coupling.  The region where this disparity is greatest is highlighted in pink.  The blue dash-dotted curves show the results of the $\calMbc = 0.0$ simulations where gas particles are staggered at half a mean particle spacing in each dimension from the dark matter particles. }  \label{fig:histograms}}
\end{figure*}

In Figure~\ref{fig:phase}, we plot temperature--density phase diagrams
for six simulations.   The top
  (bottom) panel shows a two-dimensional histograph of gas mass as
  a function of temperature and density for the GADGET (Enzo) simulations with $\calMbc
  = 0.0, 1.9,$ and $3.8$ (from left to right, respectively). The top right panel shows the impact of particle coupling.  The region
  above the dashed lines can cool in a Hubble time owing to molecular
  hydrogen (see \S \ref{ss:sf} for details), and the dotted lines show a single arbitrary adiabat ($T_g
  \propto \rho^{2/3}$).  The deviation from a single adiabat at low
  temperatures and densities is determined by Compton heating off of the
  cosmic microwave background.  Although subtle, near the mean
  density of the gas ($\rho \approx 4\times10^{-27}\,$g/cm$^3$) the
  simulations with a larger dark matter--baryon velocity difference have an $\sim 1.5$ times broader
  distribution in temperature at fixed density. 

In Figure~\ref{fig:histograms}, we show
probability distribution functions (PDFs) of the gas density and
temperature in the simulations.  For $\vbc >0$, the gas density PDFs in both the Enzo and
GADGET simulations largely agree, with some gas in the Enzo
simulations reaching higher densities than in GADGET, likely because
Enzo has higher resolution in over densities.  For $\vbc = 0$, we find
significant disagreement between the overall shape of the density PDF, as highlighted in pink in
Figure~\ref{fig:histograms}.
This disagreement owes to spurious particle coupling between
the dark matter and gas in GADGET.  \change{The amount of particle coupling is somewhat reduced when staggering the gas particles
at half an interparticle spacing in each dimension from the dark matter particles (whose locations are set by a glass file as is often done; blue dash-dotted curves) rather than using two separate glass files (the method advocated in \citealt{yoshida03}; blue dotted curves). This reduced coupling results because the dark matter and gas particles have a reduced probability of being near one another in the staggered case.  Thus, we also recommend staggering particles rather than two separate glass files.  }

The impact of particle coupling is especially apparent when comparing the peak densities that the GADGET particles reach in the different simulations.  When $\vbc >0$, the GADGET simulations appear to be less dense than Enzo at the highest densities.  For $\vbc = 0$, the simulations with GADGET have \emph{denser} gas than the Enzo simulations.  There are also significant disparities at $\delta_b \sim 10$ ($\rho_b \approx 10^{-26}~{\rm g~cm}^{-3}$) in the $\vbc = 0$ case between the GADGET and Enzo.  

\end{appendix}

\end{document}